\documentclass[fleqn,10pt]{wlscirep}
\usepackage{amsmath}
\usepackage[T1]{fontenc}
\usepackage{times}
\usepackage{lmodern}
\usepackage{amssymb}
\usepackage{bm}
\usepackage{graphicx}
\usepackage{gensymb}
\usepackage{mymacros}
\usepackage{mathtools}
\usepackage{float}
\usepackage{dsfont}
\usepackage{romannum}
\usepackage{multirow}
\usepackage{amsthm}
\usepackage{geometry}
\usepackage{lineno}
\usepackage{caption}
\usepackage{colortbl}
\usepackage{subfig}
\usepackage{xr}
\usepackage{appendix}
\usepackage{textcomp,libertine}
\usepackage{hyperref}
\newlength{\hcolwidth}
\newcommand\enuf{\cellcolor{gray}}
\newcommand\nope{\color{red}}

\title{Sparse recovery of undersampled intensity patterns for coherent diffraction imaging at high X-ray energies }

\author[1,*]{S. Maddali}
\author[1]{I. Calvo-Almazan}
\author[2]{J. Almer}
\author[2]{P.Kenesei}
\author[2]{J.-S. Park}
\author[2]{R. Harder}
\author[3,4]{Y. Nashed}
\author[1]{S. Hruszkewycz}

\affil[1]{Materials Science Division, Argonne National Laboratory, Lemont IL 60439 (USA)}
\affil[2]{X-ray Sciences Division, Argonne National Laboratory, Lemont IL 60439 (USA)}
\affil[3]{Mathematics \& Computer Science Division, Argonne National Laboratory, Lemont IL 60439 (USA)}
\affil[4]{Department of Electrical Engineering \& Computer Science, Northwestern University, Evanston IL 60208 (USA)}

\affil[*]{\textbf{Corresponding author}: smaddali@anl.gov}

\begin{abstract}
	Coherent X-ray photons with energies higher than 50 keV offer new possibilities for imaging nanoscale lattice distortions in bulk crystalline materials using Bragg peak phase retrieval methods.
However, the compression of reciprocal space at high energies typically results in poorly resolved fringes on an area detector, rendering the diffraction data unsuitable for the three-dimensional reconstruction of compact crystals. 
To address this problem, we propose a method by which to recover fine fringe detail in the scattered intensity.
This recovery is achieved in two steps: 
multiple undersampled measurements are 
made by in-plane sub-pixel motion of the area detector, then this data set is passed to a sparsity-based numerical solver that recovers fringe detail suitable for standard Bragg coherent diffraction imaging (BCDI) reconstruction methods of compact single crystals. 
The key insight of this paper is that sparsity in a BCDI data set can be enforced by recognising that the signal in the detector, though poorly resolved, is band-limited. This requires fewer in-plane detector translations for complete signal recovery,
while adhering to information theory limits.
We use simulated BCDI data sets to demonstrate the approach, outline our sparse recovery strategy, and comment on future opportunities.
 \end{abstract}
\begin{document}

\flushbottom
\maketitle
\thispagestyle{empty}
\section{Introduction}
\label{S:intro}
\begin{figure}
	\centering
	\subfloat[]{
		\includegraphics[width=0.35\hcolwidth]{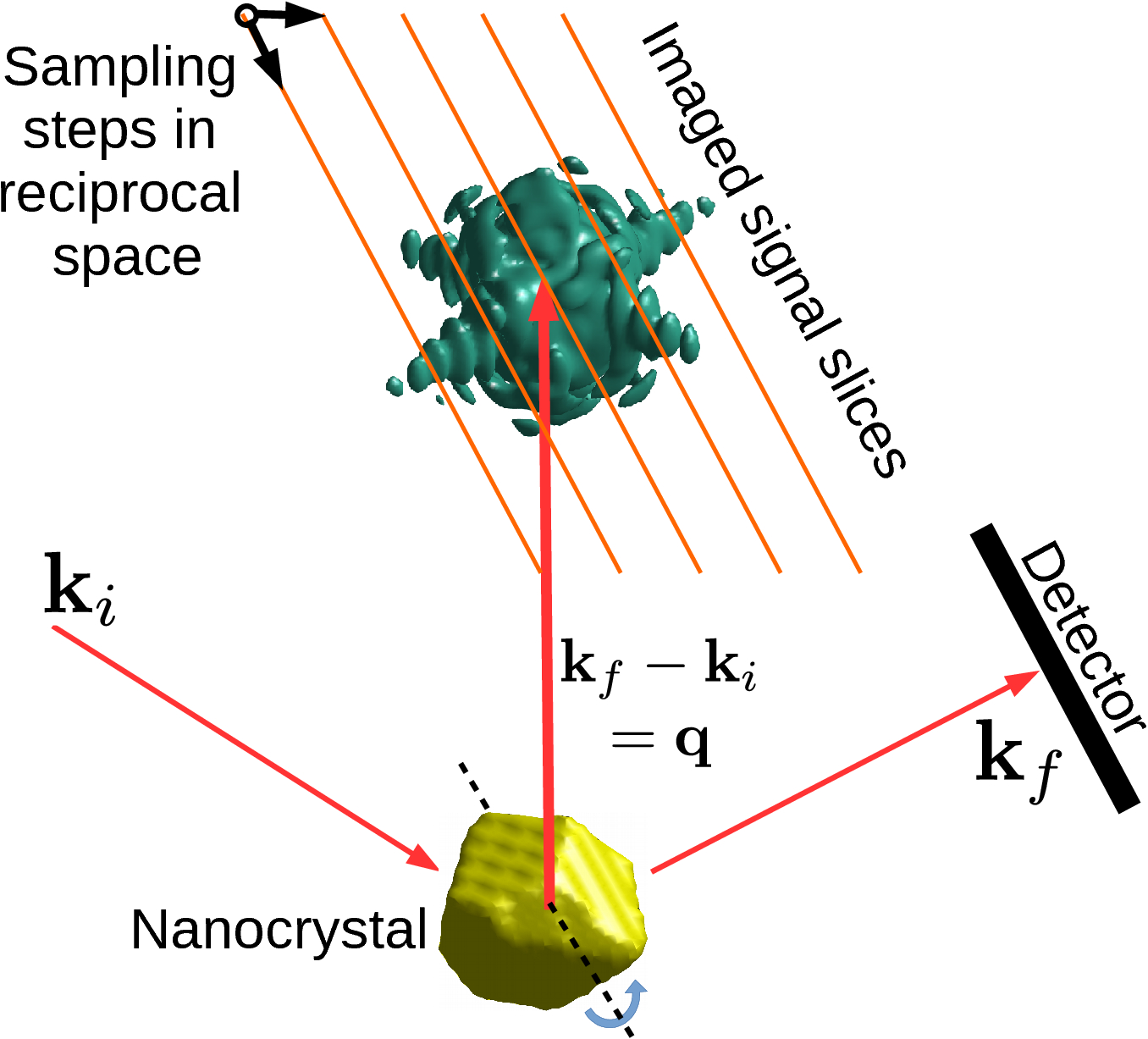}
		\label{fig:bcdi-schematic}
	} \hfill
	\subfloat[]{
		\includegraphics[width=0.55\hcolwidth]{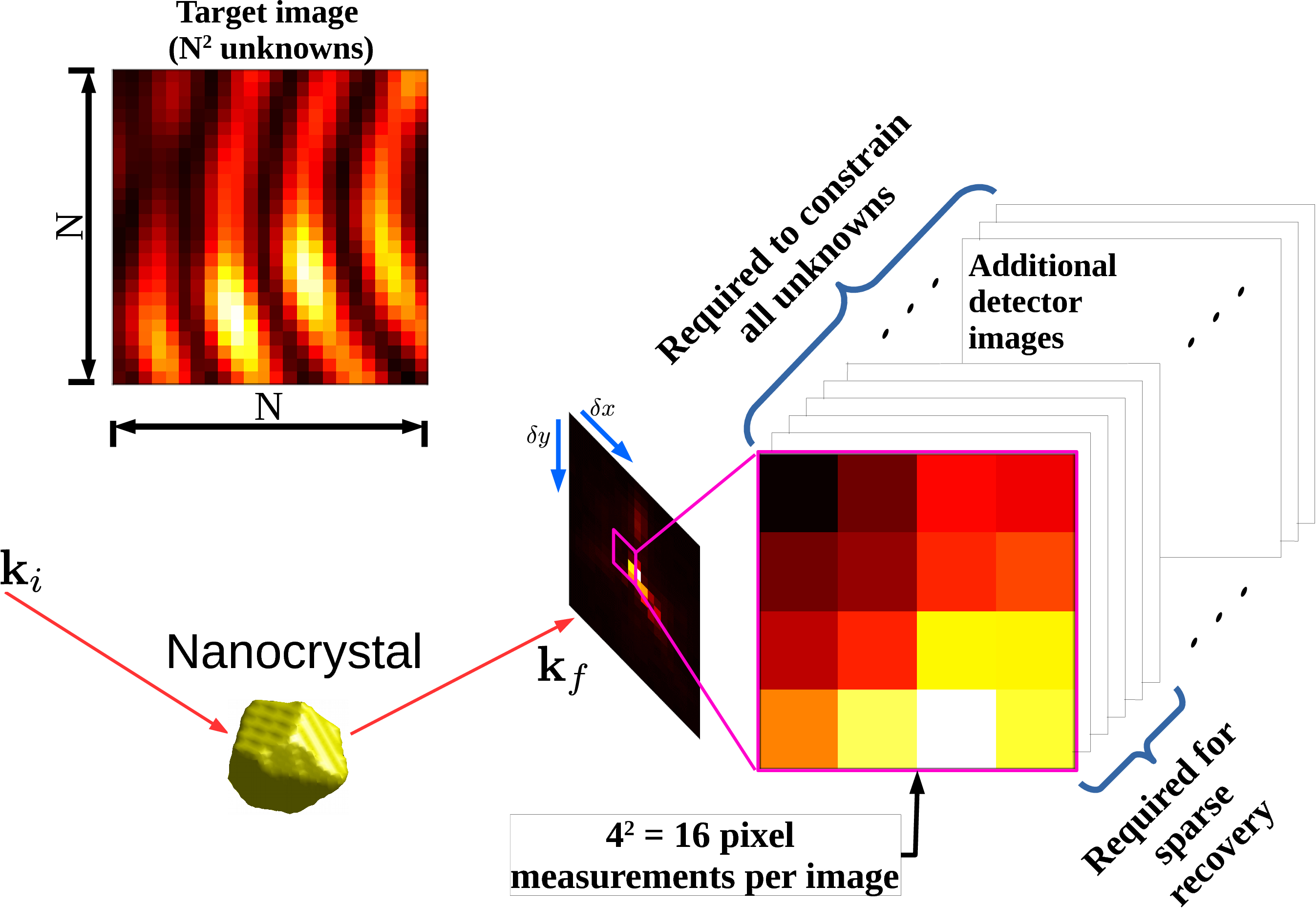}
		\label{fig:upsampling-theme}
	}
	\caption{
		\textbf{\protect\subref{fig:bcdi-schematic}} 
		Bragg CDI experimental geometry, where $\mathbf{k}_i$ and $\mathbf{k}_f$ are the incident and scattered wave vectors. 
		Three-dimensional reciprocal space is sampled in finite steps as shown. 
		Two of the three steps are in the detector plane (one of which is perpendicular to the figure plane, denoted by `$\mathbf{\circ}$') and the third dictates the migration of the plane.
		This set of sampling vectors is fixed by the the crystallographic orientation of the scatterer with respect to the beam, and the manner of its rotation.
		\textbf{\protect\subref{fig:upsampling-theme}} Acquisition of more data by detector translations perpendicular to the incoming beam in case of coarse detector resolution. Application of sparse recovery techniques greatly reduces the number of additional measurements required.
		}
	\label{fig:bcdi-upsampling}
\end{figure}

Coherent X-ray diffraction imaging (CDI) methods applied to Bragg peaks have emerged as a powerful tool in materials science for characterising lattice distortion fields and defects in crystalline nanostructures~\cite{Robinson2001,Robinson2009,Ulvestad2015,Cha2016}.
Such experiments are currently feasible at photon energies in the lower end of the hard X-ray spectrum ($9$ keV to $16$ keV), beyond which third-generation synchrotrons have very limited coherent flux.
However, ongoing and planned construction of next-generation synchrotrons will bring about greatly increased coherence at beam energies greater than $50$ keV, making high-energy Bragg CDI feasible.
This capability will enable nanoscale structural characterization of material volumes in environments accessible only with highly penetrating X-rays. 
For example, a critical new research area will be to probe deformation states within the individual grains of  macroscopic volumes, which find wide application as structural and functional materials.
Such measurements will complement and enhance models gleaned from existing non-destructive imaging techniques such as high-energy diffraction microscopy (HEDM)~\cite{Suter2008, Bernier2011} and diffraction contrast tomography~\cite{Ludwig2008} by providing structure and strain resolution within grains. 
As a step in this direction, this paper addresses prominent issues that arise when imaging individual nano- or micro-scale crystals with high-energy coherent X-rays.
In the context of structural materials, the compact crystal is a proxy for a single grain in a polycrystal. Otherwise, it can represent an isolated nanocrystal in a dense medium that requires the penetrative power of high energy X-rays (e.g. high temperature catalysis, or materials growth), opening yet more potential avenues of study~\cite{Liu2015}. 
Our treatment is focused on simulations of a BCDI measurement, in which a single compact crystal, illuminated uniformly by coherent high-energy X-rays, is rotated through the Bragg condition. 
Consequently a three-dimensional volume of reciprocal space is measured with an area detector in a sequence of parallel layers~(Figure~\ref{fig:bcdi-schematic}). 

In a BCDI experiment, increasing the beam energy from a value currently suitable for CDI (about 9 keV) to one necessary for HEDM ($\geq$ 50 keV) compresses reciprocal space by at least 80\% in each dimension. 
Under these conditions, the solid angle subtended by each detector pixel covers a substantial portion of the diffraction pattern. As a result signal features (diffraction fringes) critical to inversion of the diffraction pattern are  no longer resolved by a typical pixelated X-ray area detector. This is demonstrated in Figure \ref{fig:upsampling-theme} which shows a coherent diffraction intensity pattern depicted with well-resolved and under-resolved fringes, mimicking measurements made at lower and higher X-ray energies respectively.
The subsequent loss of diffraction feature visibility at higher energies renders conventional phase retrieval ineffective.

For the recovery of fine feature detail in the intensity pattern, additional independent measurements (upsamples) of the wave field intensity are therefore necessary. 
In this work, we consider the case where these additional measurements come from translating the area detector perpendicular to the incoming beam in sub-pixel steps, 
as has been explored previously~\cite{Chushkin2013}. 
Each pixel measurement in such a data set imposes a constraint on a particular region of the scattered intensity (Figure~\ref{fig:upsampling1}). 
In theory, the fine detail at a desired sub-pixel resolution could be recovered if one were to acquire an adequate number of these constraints, taken from sufficiently small detector offsets.
The difficulties in fulfilling this requirement are described in detail in Section~\ref{sup:underdetermined} of the Appendix.
In this paper we utilise compressed sensing~\cite{Candes2006,Donoho2006} to demonstrate signal recovery with significantly fewer of such pixel constraints than there are fine pixels in the desired image.
The required number of constraints is dictated by the information content when the desired image is expressed in Fourier space.
The reduced number of constraints (or measurements, in compressed sensing parlance) is made possible by the fact that the BCDI intensity pattern of a compact single crystal is necessarily band-limited. 
In the compressed sensing approach employed here, one needs to make approximately as many coarse pixel measurements of the poorly-resolved signal as there are non-zero components in the Fourier representation of the well-resolved signal.
The specific representation used in our method is the cosine basis. Recovery of this condensed representation of the signal is perfectly suited to mathematical optimization algorithms that specialize in sparse arrays. Throughout this paper, we refer to the upsampled intensity patterns resulting from this sparse recovery scheme as the `recovered' images or patterns.

This paper is organised in the following manner: 
Beginning with simplifying assumptions, we describe our BCDI simulations at a range of energies from 9 keV through to 54 keV, the latter corresponding to a high-energy experiment conceivable at a high-brightness synchrotron.
This includes establishing a suitable ground truth intensity pattern with well-resolved intensity fringes consistent with currently feasible measurements, against which all recovered intensity patterns are benchmarked. 
We describe the detector sub-pixel translation method and the additional constraints provided by the intrinsic band-limit of the intensity pattern, followed by the compressed sensing algorithm employed to recover each slice of the 3D data set independently. 
We demonstrate the recovery of  qualitatively different intensity patterns, with simulations that represent different beam energies and varying sub-pixel detector translations.
Finally, we show three-dimensional reconstructions of a simulated nanocrystal obtained by applying conventional BCDI phase retrieval to simulated high-energy data sets recovered with our approach.
We describe how our method can emulate a variety of physical modifications to a BCDI experiment that may be difficult to realise experimentally.
We close with comments on new experiments that this method can potentially enable at next-generation synchrotrons.
  
\section{High-energy BCDI simulations}
\label{S:HEDMsim}
BCDI measurements query the wave intensity in the Fraunhofer regime~\cite{Goodman2005} in which the scattered wave front is the Fourier transform (FT) of the illuminated compact crystal. As the scatterer is rotated through the Bragg condition the intensity pattern is measured, slice by slice, on the area detector.
For a crystal of size $\sim 300$ nm illuminated with hard X-rays of energy $\sim 9$ keV, the required range of this rotation is sufficiently small (about 0.7$^\circ$) that the successive slices are approximated as parallel in reciprocal space~\cite{Cha2016}.
\begin{figure}
	\centering
	\includegraphics[width=\hcolwidth]{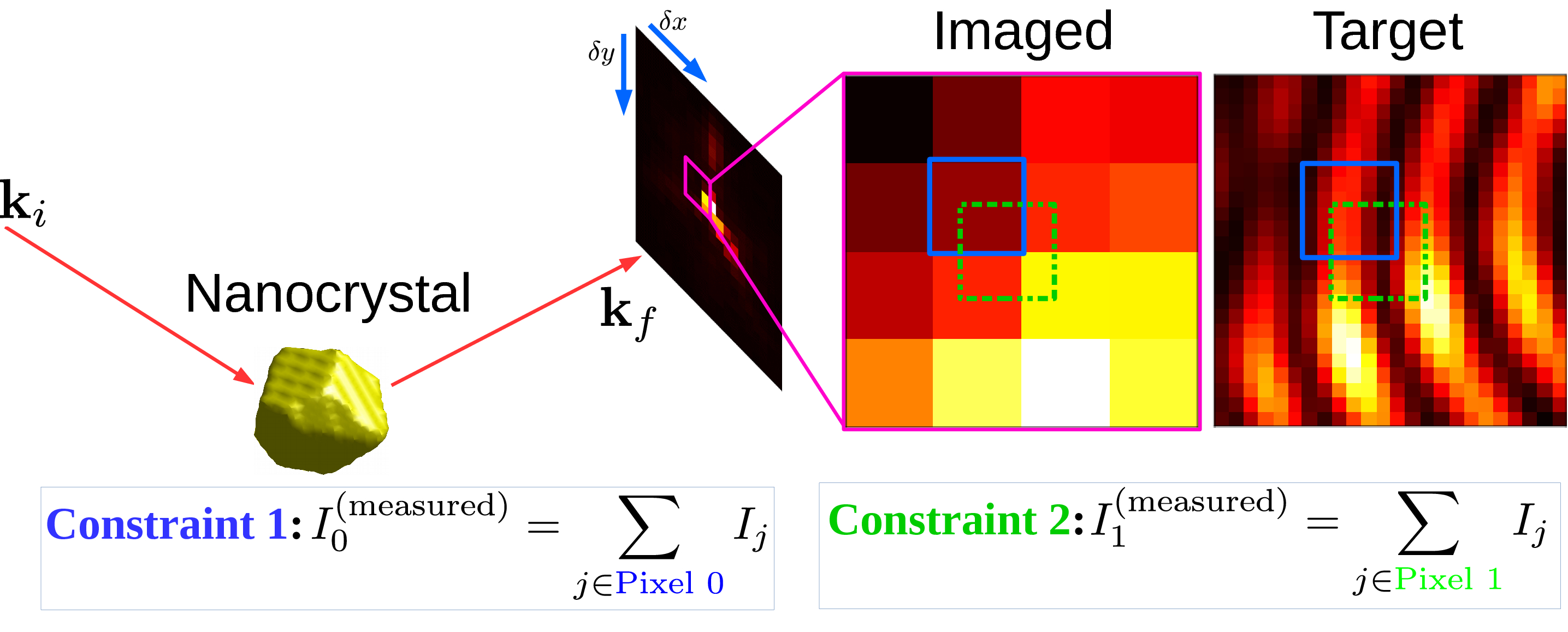}
	\caption[Upsampling strategy 1]{
		The target of the recovery algorithm is a sufficiently well-resolved approximation to the continuous scattered field. With each sub-pixel translation of the detector, new constraints on the its pixel values are obtained.
}
	\label{fig:upsampling1}
\end{figure}

We simulate the compact crystal as the set of grid points inside a faceted volume, at the centre of a three-dimensional complex array. 
The interior points are complex-valued with magnitude $1$ and a spatially varying phase that mimics a strain field. 
The exterior points have magnitude $0$. 
The corresponding diffraction signal is obtained via the three-dimensional FT. In the 3D array of this FT, two dimensions represent pixel coordinates of the area detector while the third represents successive images acquired by rotating the crystal through the Bragg condition.
In our simulations, the crystal resides inside a $22 \times 24 \times 22$ box within the simulation array of size $128\times128\times70$, denoting $70$ angular steps with a $128 \times 128$ pixel detector.
The ground truth intensity pattern is taken to be the squared modulus of the FT of this  array.
We nominally associate a beam energy of $9$ keV to this ground truth.
A data set of this kind could conceivably be collected at an existing BCDI facility. 

Simulation of an overbinned diffraction signal at higher energies is now straightforward: 
for each 2D slice in the ground truth, blocks of pixels are summed to a single intensity value (Figure~\ref{fig:upsampling1}).
This binning operation mimics a high-energy BCDI experiment since photons that would have spread over a larger solid angle at lower x-ray energies now aggregate into fewer pixels at higher energies because of the compression of reciprocal space.
Equivalently, one may imagine the overbinning to arise from a $9$ keV measurement with proportionately larger pixels. 
The ratio of pixel sizes (`pixel binning factor' of PBF) is equal to the ratio of the beam energies.
For example, the diffraction features at $9$ keV contained in every $6 \times 6$ block of pixels is squeezed into a single pixel when the beam energy is $54$ keV.
In this case, $\text{PBF} = 6$.

In this manner binning effectively reduces the feature visibility of a BCDI intensity measurement (Figure~\ref{fig:upsampling-theme}), which we redress through additional information collected from sub-pixel detector translations.
A similar method for ptychography has been proposed by Batey~\emph{et al}~\cite{Batey2014}.
 In general the finite size of the physical pixels makes it impossible to sufficiently constrain the intensity features of the ground truth that are lost to binning,
regardless of how finely the detector is translated in its plane. 
This assertion is rigorously proved in Section~\ref{sup:underdetermined} of the Supplementary Material.
 As mentioned earlier, our approach instead is to demonstrate sub-pixel-scale feature recovery with fewer measurements, which is enabled by the sparsity of the Fourier representation of the scattered intensity.
This is the focus of Section~\ref{S:compsensesparserecov} in which we discuss our binned data sets in the context of a compressed sensing measurement and motivate the use of sparse recovery techniques. 

We point out that in a real-world BCDI experiment at energies $>$ 9 keV, a given detector would span a larger $q$-space aperture that would not be entirely covered at $9$ keV.
In our simulations we consider a fixed $q$-space aperture corresponding to the $128\times128$ detector space of the ground truth.
We focus on the recovery of fine features in the original ground truth simulation from binned data sets that subtend this $q$ range.
We justify this by pointing out that in typical BCDI experiments at $9$ keV the edge pixels of the detector capture relatively low intensities compared to the central pixels (by a few orders of magnitude).
The scattered intensities outside this aperture contribute negligibly to the Fourier representation of the original scatterer.

\section{Sparse recovery: mathematical details}
\label{S:compsensesparserecov}
\begin{figure}
	\centering
	\subfloat{\includegraphics[width=0.3\hcolwidth]{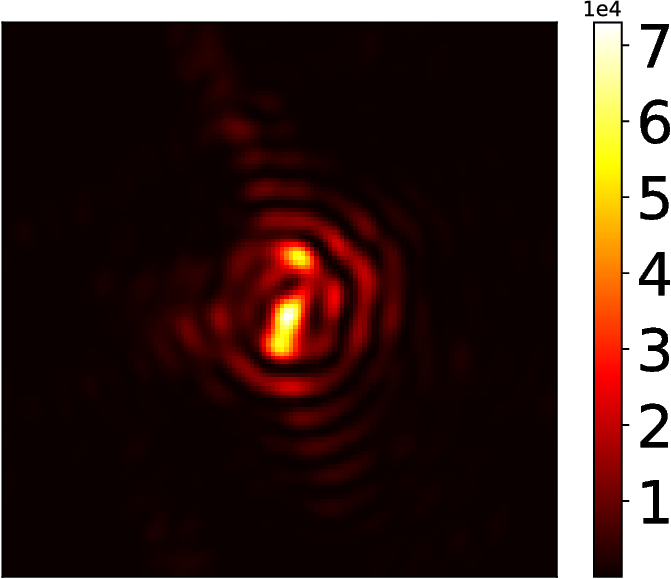}\label{fig:sparseBasis-1}} \hfill
	\subfloat{\includegraphics[width=0.3\hcolwidth]{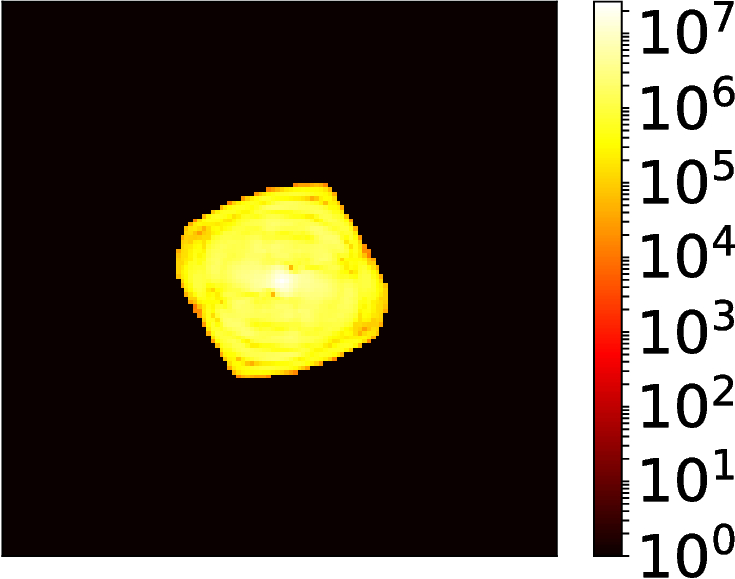}\label{fig:sparseBasis-2}} \hfill
	\subfloat{\includegraphics[width=0.3\hcolwidth]{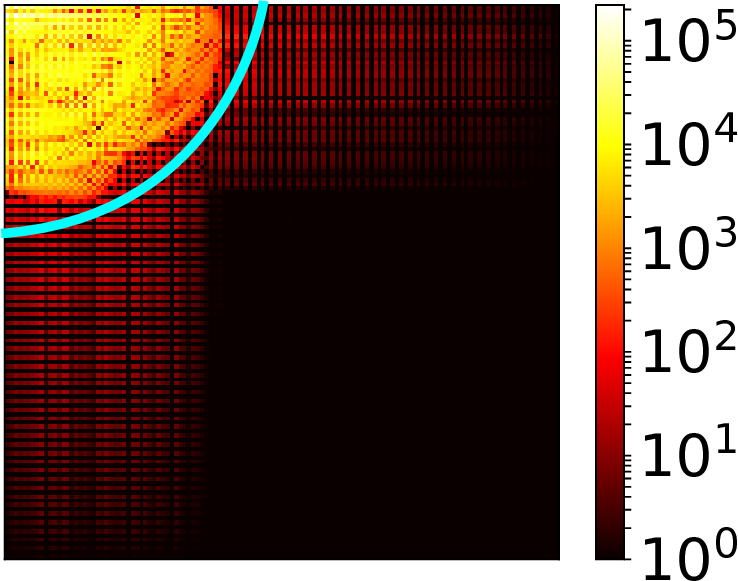}\label{fig:sparseBasis-3}} \vfill
	\caption{
		\textbf{\protect\subref{fig:sparseBasis-1}} Diffraction from a faceted object,
		\textbf{\protect\subref{fig:sparseBasis-2}} Magnitude of its FT, 
		\textbf{\protect\subref{fig:sparseBasis-3}} Magnitude of its 2-D DCT.
		In the cosine basis, the spectral components inside the marked region contribute the most to \textbf{\protect\subref{fig:sparseBasis-1}}.
		}
	\label{fig:sparseBasis}
\end{figure}

We seek a method to reverse the binning process described in Section~\ref{S:HEDMsim} and obtain the original ground-truth diffraction pattern that features well-resolved fringes.
In our method, recovery of the fine detail from a limited set of binned pixel measurements hinges on representing each two-dimensional slice in the discrete cosine basis, which we utilise as a numerically convenient variant of the FT.
The two-dimensional discrete cosine transform (DCT) is defined for an $N \times N$-sized image $A_{ij}$ as the linear transformation ~\cite{Makhoul1980}:
\begin{equation}
	\left[\text{DCT}\left(A_{ij}\right) \right]_{mn}
	= \sum_{i=0}^{N-1} \sum_{j=0}^{N-1}
	A_{ij} 
	\mathds{C}(i;m) \mathds{C}(j;n)
	\label{eq:DCT2}
\end{equation}
where $\mathds{C}(p;q) = \cos\left[\left(p+1/2\right)q\pi/N\right]$. 
This transform is instrumental in many digital media compression formats such as JPEG and MP3~\cite{Wallace1991,Musmann2006}. 
Figures~\ref{fig:sparseBasis-2} and~\ref{fig:sparseBasis-3} compare the FT and DCT representations of the diffraction pattern in~\ref{fig:sparseBasis-1}. 
We demonstrate in Section~\ref{S:results} that the results of the sparse recovery algorithm do not depend on the smallest DCT components.
Thus, our simulations do not consider detector noise because for a high signal-to-noise ratio ($\simeq 500$ dB), the noise manifests weakly in the DCT and is automatically removed by the optimisation.

The three-dimensional diffraction pattern from a finite crystal is inherently a band-limited function, since its FT is the auto-correlation of the scattering factor of a finite crystal (the Patterson function, Figure~\ref{fig:sparseBasis-2}). 
From this it follows that each two-dimensional slice of the diffraction pattern is also band-limited (this is proved rigorously in Section~\ref{sup:appA} of the Appendix). 
Since each slice contains essentially the same signal information as its two-dimensional DCT, one would require approximately as many pixel measurements as the number of non-zero components in this latter representation in order to fully determine it.
This well-known principle from information theory underlies all compressed sensing/sparse recovery algorithms. 
This approach is of immense importance in scenarios where a scant number of physical measurements seemingly fail to mathematically constrain a system.
After sparse recovery, the target intensity pattern we ultimately seek is obtained simply by inverting this transform. 

To avoid information loss in our ground truth simulation, the Patterson function of the simulated crystal has to be fully captured within the three dimensional simulation array.
In other words, the buffer width (populated with zero-valued pixels) around the scatterer should be at least the span of the scatterer in each dimension (this is equivalent to the Nyquist sampling criterion~\cite{Shannon1949}).
The simulation sizes described in Section~\ref{S:HEDMsim} go well above this minimum requirement to ensure that the Patterson function is not only fully captured, but also sparse in the array. 

We define the sparse recovery problem in terms of (\romannum{1}) The unknown (target) diffraction slice $\mathbf{I}$ of the desired fine-pixel resolution, which is sparse in the DCT representation: $\mathbf{I} \equiv B\mathbf{x}$ (where $\mathbf{x}$ is sparse and the columns of matrix $B$ are the inverse DCT basis vectors) and (\romannum{2}) a set of measurements made on this signal, represented by a linear operation $A$ resulting in measured (binned) values $\mathbf{I}^{\text{(measured)}} = A\mathbf{I} = A\left(B\mathbf{x}\right)$. 
We illustrate these operations using the example intensity pattern shown in Figure~\ref{fig:upsampling1} that has a relatively small number of pixels.
In this example,  the coarse pixelation of a fine grid of size $28 \times 28 = 784$ pixels results in a total of $4 \times 4 = 16$ measurements, and therefore $A$ is a $16 \times 784$ matrix.
Additional measurements are obtained by translating the detector in the plane perpendicular to the incoming beam as shown.
Each detector offset provides a new set of pixel measurements (fewer than $16$ per detector translation, since we ignore the pixels that fall outside the original $q$ range of interest).
Each pixel measurement corresponds to a row of the matrix $A$.
As we have proved in Section~\ref{sup:underdetermined} of the Appendix, it is not possible to obtain $784$ independent rows of the matrix $A$ through detector shifts, and we address this problem with compressed sensing.
All compressed sensing techniques solve the system of equations $AB\mathbf{x} = \mathbf{I}^{(\text{measured})}$ by enforcing that $\mathbf{x}$ be sparse.

While there exist various algorithms for sparse recovery~\cite{Suykens2000,Huang2008,Candes2011}, we adopt the LASSO regression method~\cite{Tibshirani1996} common in machine learning applications:
\begin{equation}
	\mathbf{x}_{\text{optimal}} = \arg \min_{\mathbf{x}} 
	\left\{
		\left|AB\mathbf{x} - \mathbf{I}^{\text{(measured)}}\right|^2 + 
		\alpha \left|\mathbf{x}\right|
	\right\}
	\label{eq:LASSO}
\end{equation}
for some small $\alpha > 0$ (set to $2\e{-4}$ throughout this paper).
The $\left|\mathbf{x}\right|$ penalty imposition on the objective function in Equation~\eqref{eq:LASSO} explicitly enforces sparsity on the unknown $\mathbf{x}$ ($\ell_1$-optimisation). 
Briefly, the optimisation converges to that solution $\mathbf{x}$ which has the fewest non-zero components and simultaneously satisfies the set of constraints $AB\mathbf{x} = \mathbf{I}^{\text{(measured)}}$.
The target image is recovered by inverting the sparsifying transform ($\mathbf{I} = B\mathbf{x}_{\text{optimal}}$). 
We point out that this recovery scheme does not strictly enforce the constraint of non-negativity on $\mathbf{I}$.
This results in recovered images with a few negative-valued pixels, which we simply threshold to zero (Figures~\ref{fig:origImageLog},~\ref{fig:recovImageLog}).
We show in Section~\ref{S:results} that this thresholding still results in a working approximation of the continuous signal, from a phase retrieval point of view.

Compressed sensing theory provides a more precise success threshold for the number of pixel measurements: for an image of size $N \times N$ pixels, $M \geq K \log (N^2 / K )$ where $K$ is the number of significant (nonzero) components in the sparse representation~\cite{Donoho2006}.
In general, results of the recovery when $M$ is well above this threshold are negligibly different from the original image itself.

\section{Recovery results}
\label{S:results}
\begin{figure}
	\centering
	\subfloat[]{\includegraphics[width=0.3\hcolwidth]{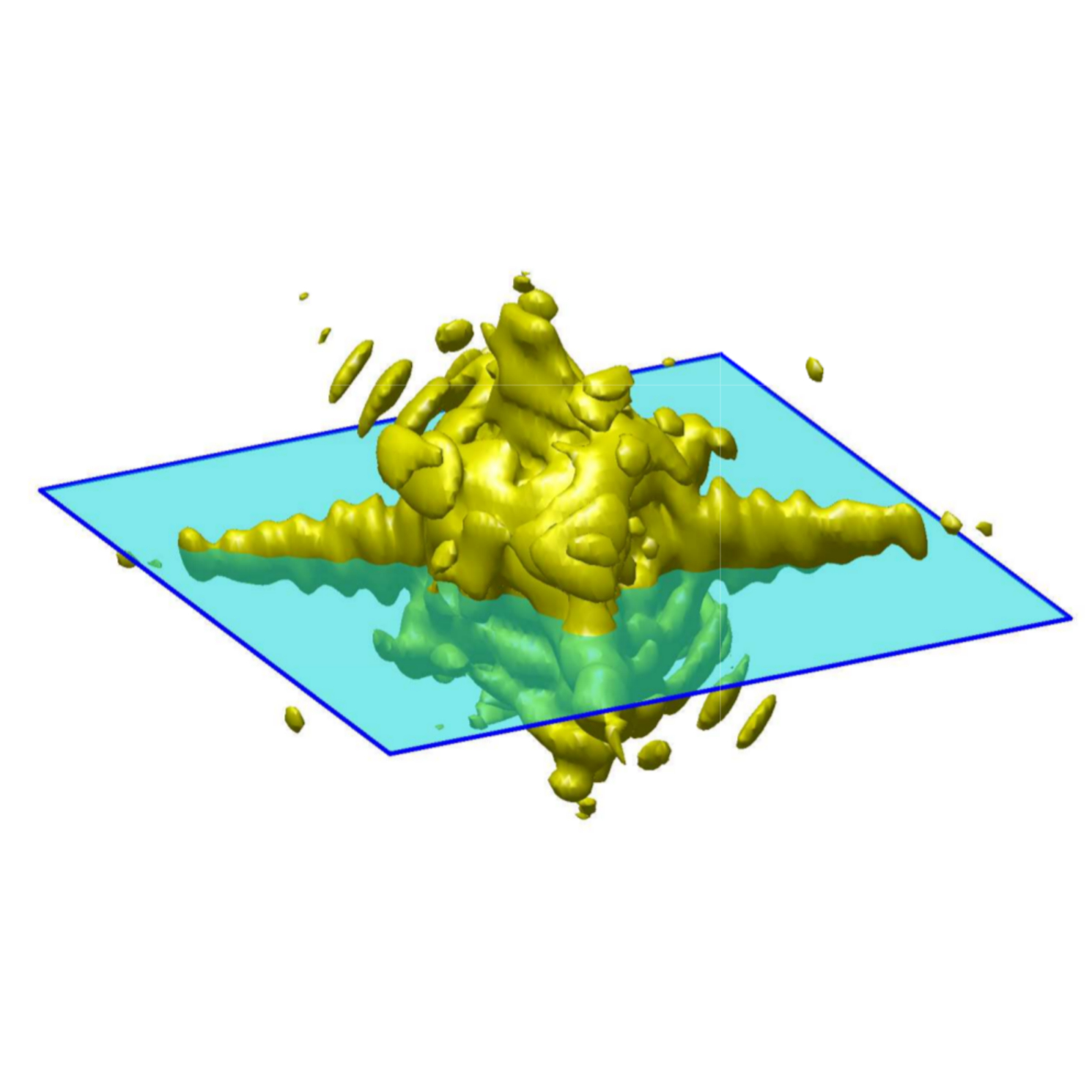}\label{fig:diffSignal}}	\hfill	
	\subfloat[]{\includegraphics[width=0.3\hcolwidth]{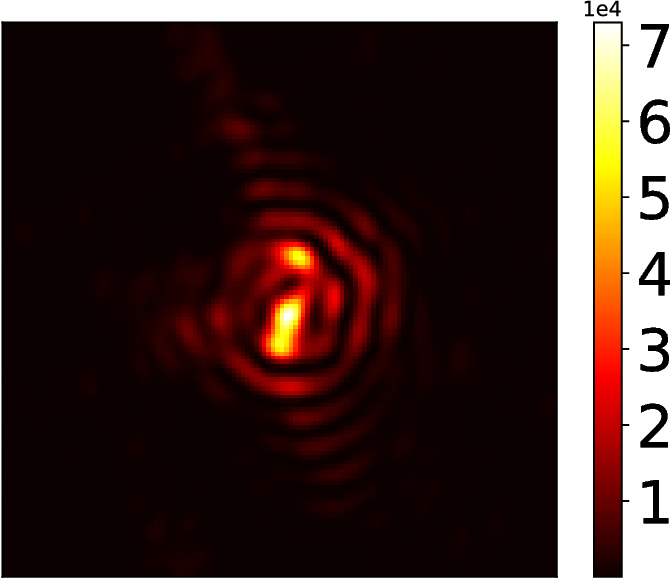}\label{fig:origImage}}	\hfill
	\subfloat[]{\includegraphics[width=0.3\hcolwidth]{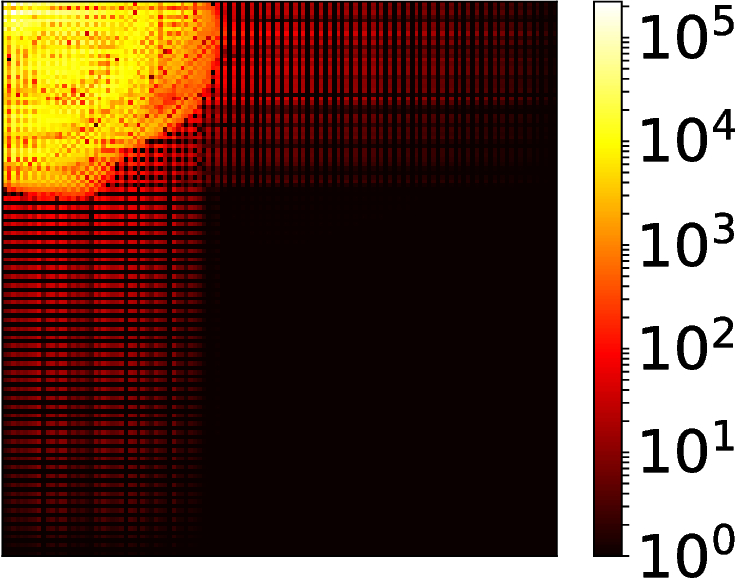}\label{fig:origSpec}} \hfill	
	\vfill
	\subfloat[]{\includegraphics[width=0.3\hcolwidth]{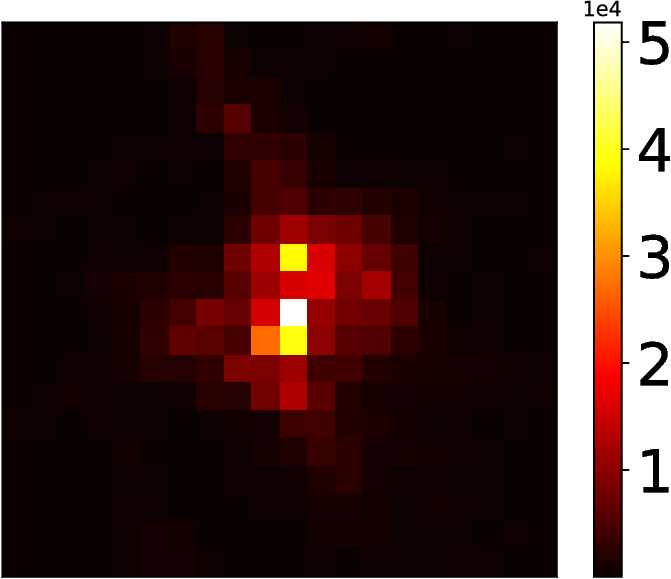}\label{fig:measuredImage}}	\hfill
	\subfloat[]{\includegraphics[width=0.3\hcolwidth]{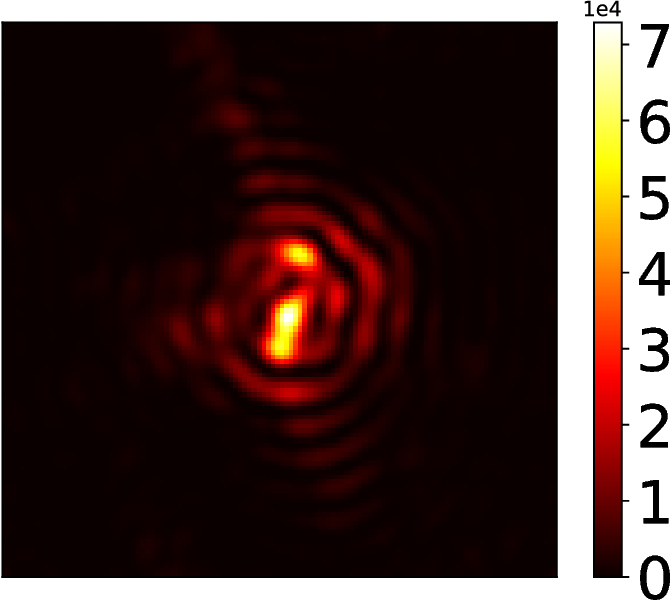}\label{fig:recovImage}}	 \hfill
	\subfloat[]{\includegraphics[width=0.3\hcolwidth]{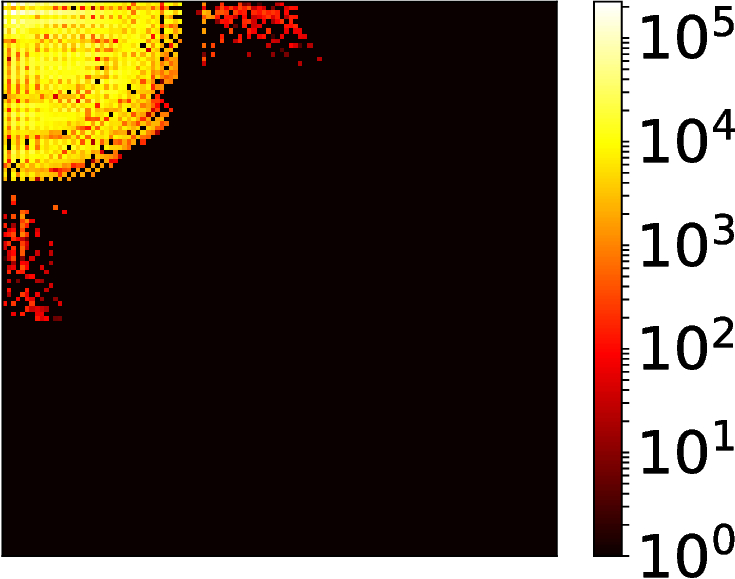}\label{fig:recovSpectrum}}	\hfill
	\vfill
	\subfloat[]{\includegraphics[width=0.45\hcolwidth]{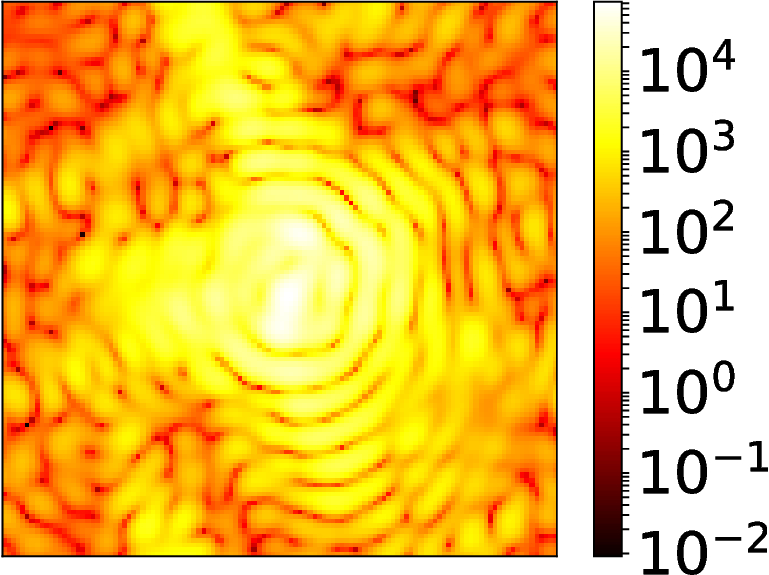}\label{fig:origImageLog}}	\hfill
	\subfloat[]{\includegraphics[width=0.45\hcolwidth]{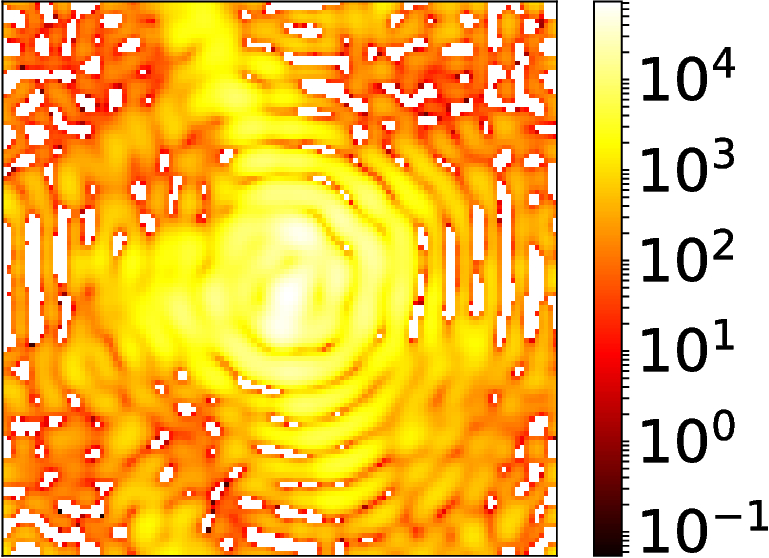}\label{fig:recovImageLog}}	\hfill
	\vfill
	\caption{
		\textbf{\protect\subref{fig:diffSignal}} Slice of 3-D diffraction signal,
		\textbf{\protect\subref{fig:origImage}} Slice imaged on detector, nominally 9 keV, 
		\textbf{\protect\subref{fig:origSpec}} DCT spectrum magnitude of 9 keV signal, 
		\textbf{\protect\subref{fig:measuredImage}} Equivalent signal in \protect\subref{fig:origImage} simulated at 54 keV, 
		\textbf{\protect\subref{fig:recovImage}} Recovered image with pixelation equivalent to 9 keV signal,
		\textbf{\protect\subref{fig:recovSpectrum}} DCT spectrum magnitude of recovered signal,
		\textbf{\protect\subref{fig:origImageLog}} Benchmark image on a log scale, and
		\textbf{\protect\subref{fig:recovImageLog}} Recovered image on a log scale, showing the regions where the negative pixels were thresholded to zero.
		}
	\label{fig:results}
\end{figure}

Figure~\ref{fig:results} shows a visual comparison between a benchmark ground truth diffraction slice ~\eqref{fig:origImage}, and the corresponding fine-pixel diffraction slice recovered from a set of simulated intensity patterns at 54 keV obtained with sub-pixel detector translation~\eqref{fig:recovImage}.
The coarsely binned diffraction slice has size $20\times20$ pixels~\eqref{fig:measuredImage}, while the recovered diffraction slice has size $120\times120$ (fine) pixels, corresponding to a PBF of 6.

\begin{figure}
	\centering
	\subfloat[]{\includegraphics[width=0.45\hcolwidth]{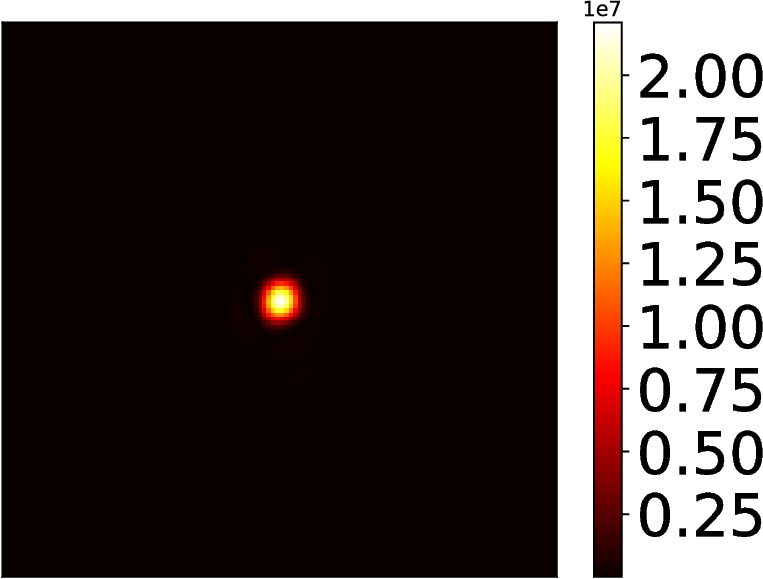}\label{fig:image_onBragg}}	\hfill			
	\subfloat[]{\includegraphics[width=0.45\hcolwidth]{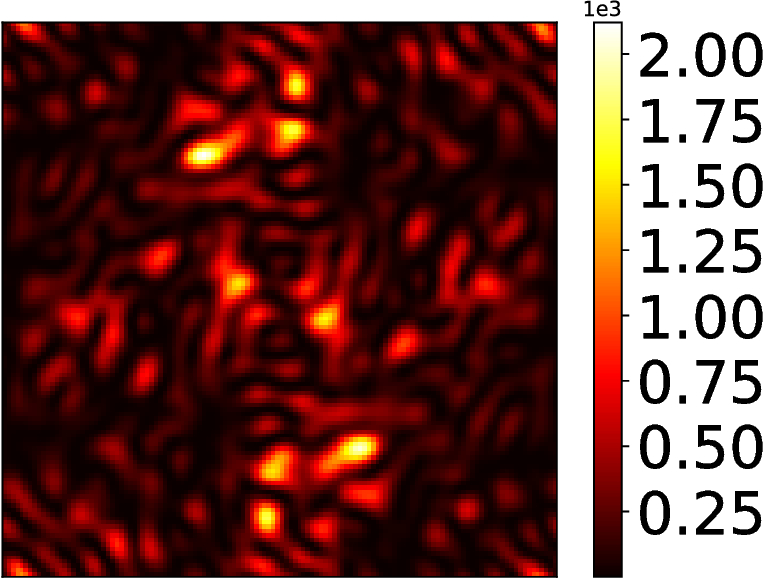}\label{fig:image_offBragg}}	\hfill	
	\vfill
	\subfloat[]{\includegraphics[width=0.45\hcolwidth]{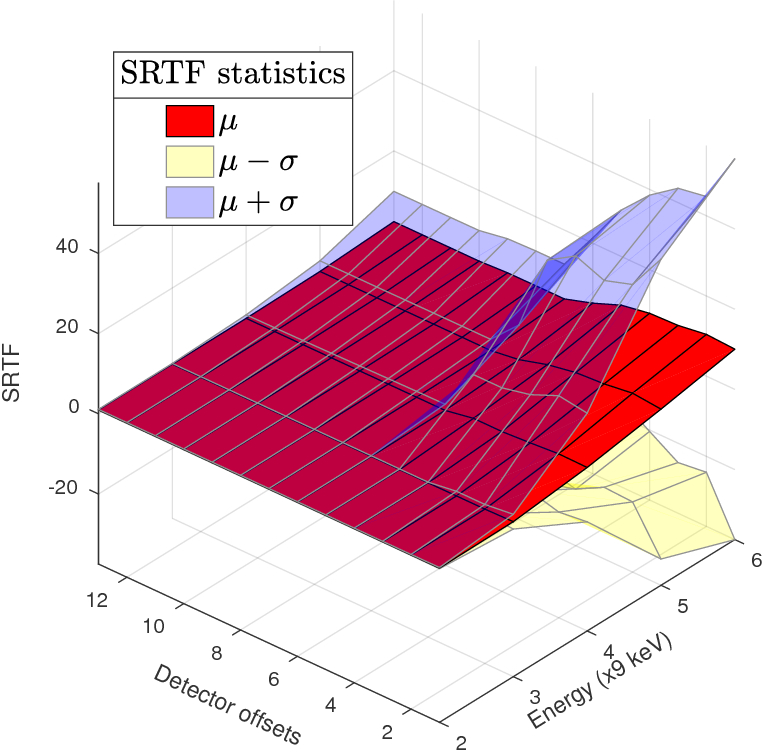}\label{fig:error_onB}}	\hfill	
	\subfloat[]{\includegraphics[width=0.45\hcolwidth]{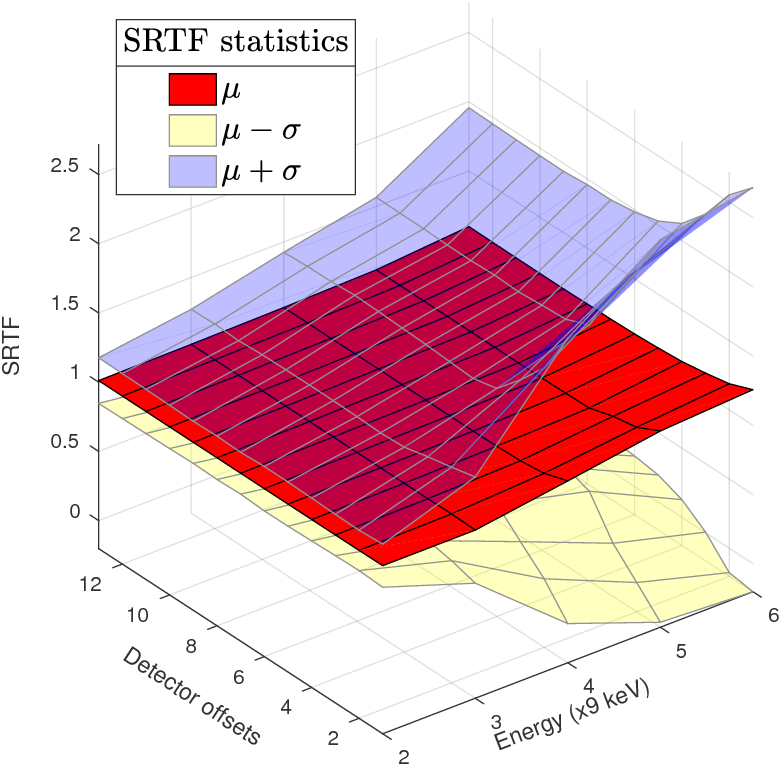}\label{fig:error_offB}} \hfill
	\vfill
	\fbox{\subfloat[]{\includegraphics[width=0.21\hcolwidth]{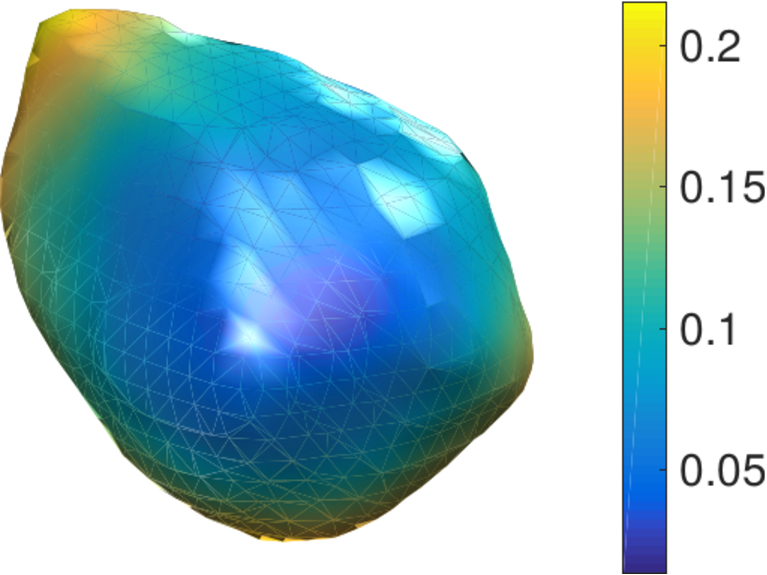}\label{fig:original}}} \hfill
	\subfloat[]{\includegraphics[width=0.21\hcolwidth]{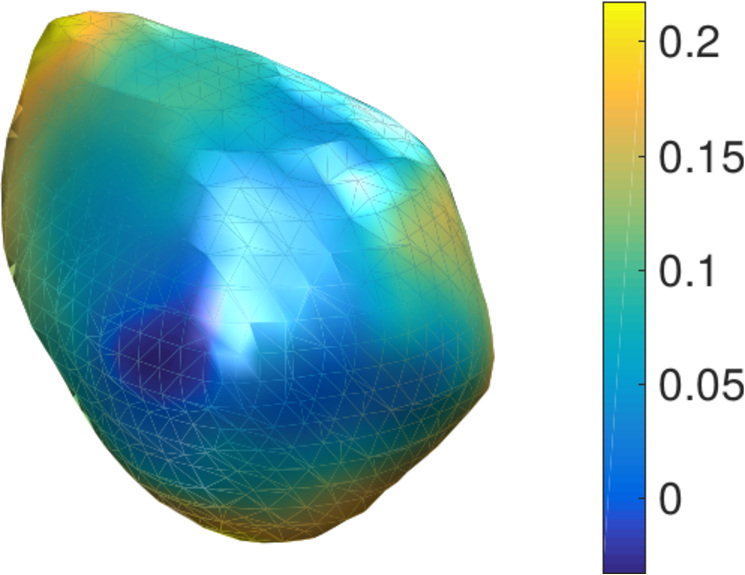}\label{fig:FPL_4}} \hfill
	\subfloat[]{\includegraphics[width=0.21\hcolwidth]{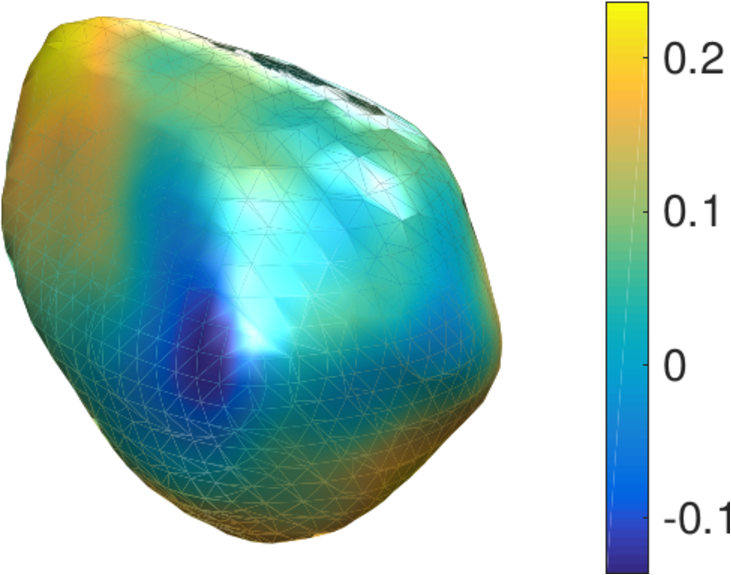}\label{fig:FPL_5}} \hfill
	\subfloat[]{\includegraphics[width=0.21\hcolwidth]{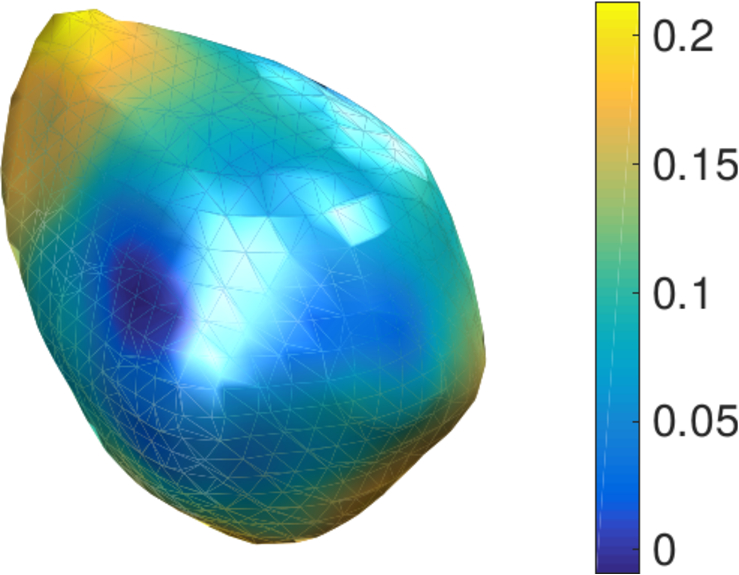}\label{fig:FPL_6}} \hfill
	\caption{
		\textbf{\protect\subref{fig:image_onBragg}} Central slice of a 3-D diffraction pattern,
		\textbf{\protect\subref{fig:image_offBragg}} edge slice of the same 3-D diffraction pattern, 
		\textbf{\protect\subref{fig:error_onB}} SRTF trend for central slice, 
		\textbf{\protect\subref{fig:error_offB}} SRTF trend for edge slice, 
		\textbf{\protect\subref{fig:original}} original synthetic particle with surface phase variation, 
		\textbf{\protect\subref{fig:FPL_4} \textemdash \protect\subref{fig:FPL_6}} reconstructed particle and surface phase variation (in radians) corresponding to energies 36 keV, 45 keV and 54 keV respectively. The phase retrieval recipe used was: solvent flipping (400 iter.)$\rightarrow$hybrid input-output ($\beta=0.8$, 240 iter.)$\rightarrow$solvent flipping (400 iter.)$\rightarrow$error reduction (100 iter.)~\cite{Marchesini2007}, with shrinkwrapping every 25 iterations.
	}
	\label{fig:errortrend}
\end{figure}

For a better quantitative picture, we examine the fidelity of two different recovered diffraction patterns to their respective 9 keV ground truth benchmark slices, as a function of the beam energy and degree of upsampling. 
The benchmark images represent qualitatively different intensity distributions:
the "on-Bragg" central slice (Figure~\ref{fig:image_onBragg}) has one strong, highly localised peak while the intensity distribution in the "off-Bragg" terminal slice (Figure~\ref{fig:image_offBragg}) is weaker and more spread out. 
The fidelity is quantified by the sparse recovery transfer function (SRTF), which we define for a single recovered image as
\begin{equation}
	\text{SRTF}(i,j) = \sqrt{
		\frac{I_\text{recovered}(i,j)}{I_\text{ground truth}(i,j)}
		}
	\label{eq:prtf}
\end{equation}
where the indices $(i,j)$ run over the pixels of the finely pixelated diffraction pattern. 
The SRTF is analogous to the phase retrieval transfer function common in phase retrieval literature~\cite{Chapman2006}.
For a perfect recovery, SRTF $=1$ for all pixels. 
Our results are expressed in terms of the mean $\mu$ of the SRTF and the standard deviation spread $\mu \pm \sigma$ around the mean, evaluated over the recovered (upsampled) images. 

\begin{figure}
	\subfloat[]{\includegraphics[width=0.45\hcolwidth]{image_onBragg}\label{fig:example}} \hfill
	\subfloat[]{\includegraphics[width=0.45\hcolwidth]{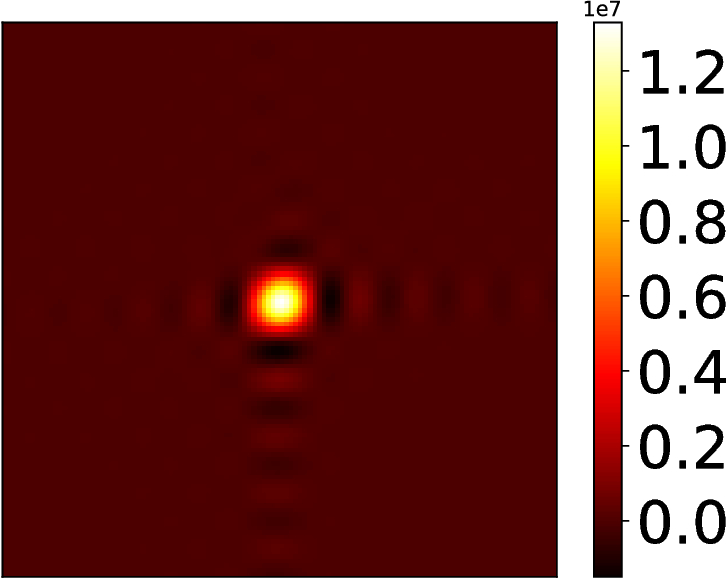}\label{fig:badRecon}}
	\caption{
		\textbf{\protect\subref{fig:example}} Central "on-Bragg" benchmark image, 
		\textbf{\protect\subref{fig:badRecon}} Bad recovery at 54 keV from a single detector position, or equivalently $20\times20 = 400$ measurements \emph{i.e.} no detector displacement.
	}
	\label{fig:badrecon}
\end{figure}

Figures~\ref{fig:error_onB} and~\ref{fig:error_offB} indicate a steady improvement in the SRTF with more detector offsets. 
Sufficient sub-pixel translations of the detector ensure that the SRTF of the recovered image remains in the vicinity of the ideal value of $1$. 
With higher beam energy, there is a greater variance in the SRTF.
This is due to the appearance of artefacts from an insufficiently constrained optimisation, as can most prominently be seen in Figures~\ref{fig:example} and \ref{fig:badRecon}.
In figures~\ref{fig:error_onB} and~\ref{fig:error_offB}, on the other hand, the slight deviation of the SRTF from $1$ at high PBF stems from the small penalty $\alpha \left|\mathbf{x}\right|$ on the objective function in Equation~\eqref{eq:LASSO}.

Finally in Figure~\ref{fig:FPL_4}~\textemdash~\ref{fig:FPL_6}  we show the results of the standard BCDI phase retrieval applied to diffraction intensity patterns upsampled from simulations of 36, 45 and 54 keV data sets (PBF $= 4, 5, 6$) with sub-pixel translations. 
Each upsampled intensity pattern data set was recovered  from the original detector position and $12$ additional detector offsets along the pixel diagonals. The reconstructions are in agreement with the original structure~\ref{fig:original}.

\section{Discussion}
\label{S:discussion}
We have outlined a framework by which fine features in high-energy coherent diffraction intensity patterns 
from individual compact crystals can be recovered using compressed sensing and sparse recovery.
The success of these methods is often justified in literature by demonstrating recovery with fewer measurements than the size of the actual signal array. 
Thus far we have also described our methodology in terms of these coarse `pixel measurements' when quantifying information content in a signal. 
In this section we translate this phrasing into quantities that are directly related to the design of CDI experiments. 
Specifically, we show how our sparse recovery technique can be exploited to emulate smaller pixel sizes or larger sample-detector distances.
We thereby demonstrate the advantage of putting the burden of signal detection on numerical algorithms, as opposed to substantially modifying an existing experimental setup or constructing massive experimental enclosures to enable large detector distances at large Bragg angles.
\begin{table}	\fontsize{7.5}{10.2}\selectfont
	\centering
	\subfloat[][]{\begin{tabular}{|c|c|c|c|c|c|c|}
\hline
	\multicolumn{ 2}{|c|}{} & \multicolumn{ 5}{c|}{\textbf{PBF (energy multiplier)}} \\ \cline{ 3- 7}
	\multicolumn{ 2}{|c|}{} & \textbf{2} & \textbf{3} & \textbf{4} & \textbf{5} & \textbf{6} \\ \hline
	\multirow{13}{*}{\rot{\textbf{Number of detector positions}}} & 
							\textbf{13} &	\enuf{14043} &	\enuf{7684} &	\enuf{6787} &	\enuf{4808} &	\enuf{4371}	\\	\cline{2-7}
	\multicolumn{1}{|c|}{} & \textbf{12} &	\enuf{14043} &	\enuf{7684} &	\enuf{6787} &	\enuf{4808} &	{4371}	\\	\cline{2-7}
	\multicolumn{1}{|c|}{} & \textbf{11} &	\enuf{14043} &	\enuf{7684} &	\enuf{6787} &	\enuf{4808} &	{4010}	\\	\cline{2-7}
	\multicolumn{1}{|c|}{} & \textbf{10} &	\enuf{14043} &	\enuf{7684} &	\enuf{6787} &	\enuf{4808} &	{3649}	\\	\cline{2-7}
	\multicolumn{1}{|c|}{} & \textbf{9} &	\enuf{14043} &	\enuf{7684} &	\enuf{6787} &	{4808} &	{3288}	\\	\cline{2-7}
	\multicolumn{1}{|c|}{} & \textbf{8} &	\enuf{14043} &	\enuf{7684} &	{6787} &	{4279} &	{2927}	\\	\cline{2-7}
	\multicolumn{1}{|c|}{} & \textbf{7} &	\enuf{14043} &	\enuf{7684} &	{5946} &	{3750} &	{2566}	\\	\cline{2-7}
	\multicolumn{1}{|c|}{} & \textbf{6} &	\enuf{14043} &	\enuf{7684} &	{5105} &	{3221} &	{2205}	\\	\cline{2-7}
	\multicolumn{1}{|c|}{} & \textbf{5} &	\enuf{14043} &	{7684} &	{4264} &	{2692} &	{1844}	\\	\cline{2-7}
	\multicolumn{1}{|c|}{} & \textbf{4} &	{14043} &	{6163} &	{3423} &	{2163} &	{1483}	\\	\cline{2-7}
	\multicolumn{1}{|c|}{} & \textbf{3} &	{10562} &	{4642} &	{2582} &	{1634} &	\nope{1122}	\\	\cline{2-7}
	\multicolumn{1}{|c|}{} & \textbf{2} &	{7081} &	{3121} &	{1741} &	\nope{1105} &	\nope{761}	\\	\cline{2-7}
	\multicolumn{1}{|c|}{} & \textbf{1} &	{3600} &	{1600} &	\nope{900} &	\nope{576} &	\nope{400}	\\	\hline
\end{tabular}
	\label{tab:canabletoRecover}

 }
	\qquad
	\subfloat[][]{	\begin{tabular}{|c|c|c|c|c|c|c|}
	\hline
	\multicolumn{ 2}{|c|}{} & \multicolumn{ 5}{c|}{\textbf{Beam energy ($\bs{\times 9}$ keV)}} \\ \cline{ 3- 7}
	\multicolumn{ 2}{|c|}{} & \textbf{2} & \textbf{3} & \textbf{4} & \textbf{5} & \textbf{6} \\ \hline
	\multirow{13}{*}{\rot{\textbf{Number of detector positions}}} & 
		\textbf{13}	& \enuf{1.975}	& \enuf{2.191}	& \enuf{2.746}	& \enuf{2.889}	& \enuf{3.306}	\\ \cline{2-7}
		\multicolumn{1}{|c|}{} & 		\textbf{12}	& \enuf{1.975}	& \enuf{2.191}	& \enuf{2.746}	& \enuf{2.889}	& 3.306	\\ \cline{2-7}
		\multicolumn{1}{|c|}{} & 		\textbf{11}	& \enuf{1.975}	& \enuf{2.191}	& \enuf{2.746}	& \enuf{2.889}	& 3.166	\\ \cline{2-7}
		\multicolumn{1}{|c|}{} & 		\textbf{10}	& \enuf{1.975}	& \enuf{2.191}	& \enuf{2.746}	& \enuf{2.889}	& 3.02	\\ \cline{2-7}
		\multicolumn{1}{|c|}{} & 		\textbf{9}	& \enuf{1.975}	& \enuf{2.191}	& \enuf{2.746}	& 2.889	& 2.867	\\ \cline{2-7}
		\multicolumn{1}{|c|}{} & 		\textbf{8}	& \enuf{1.975}	& \enuf{2.191}	& 2.746	& 2.726	& 2.705	\\ \cline{2-7}
		\multicolumn{1}{|c|}{} & 		\textbf{7}	& \enuf{1.975}	& \enuf{2.191}	& 2.57	& 2.552	& 2.533	\\ \cline{2-7}
		\multicolumn{1}{|c|}{} & 		\textbf{6}	& \enuf{1.975}	& \enuf{2.191}	& 2.382	& 2.365	& 2.348	\\ \cline{2-7}
		\multicolumn{1}{|c|}{} & 		\textbf{5}	& \enuf{1.975}	& 2.191	& 2.177	& 2.162	& 2.147	\\ \cline{2-7}
		\multicolumn{1}{|c|}{} & 		\textbf{4}	& 1.975	& 1.963	& 1.95	& 1.938	& 1.925	\\ \cline{2-7}
		\multicolumn{1}{|c|}{} & 		\textbf{3}	& 1.713	& 1.703	& 1.694	& 1.684	& \nope{1.675}	\\ \cline{2-7}
		\multicolumn{1}{|c|}{} & 		\textbf{2}	& 1.402	& 1.397	& 1.391	& \nope{1.385}	& \nope{1.379}	\\ \cline{2-7}
		\multicolumn{1}{|c|}{} & 		\textbf{1}	& 1	& 1	& \nope{1}	& \nope{1}	& \nope{1}	\\ \hline
	\end{tabular}
	\label{tab:canabletoRecover-2}
 }
	\caption{
		\textbf{\protect\subref{tab:canabletoRecover}} Number of unique binning constraints $M$ resulting from diagonal detector offsets alone, as a function of PBF.
 		Values of $M$ below the information theoretic limit are shown in red.
 		The shaded numbers indicate when further detector offsets do not contribute any new binning constraints.
		\textbf{\protect\subref{tab:canabletoRecover-2}} Effective multiplier for sample-detector distance and its variation with beam energy and number of detector offsets.
 		The colour codes are the same as for \protect\subref{tab:canabletoRecover}.
		For reference, in an earlier  BCDI experiment at $9$ keV, a $300$ nm crystal has been imaged at a sample detector distance of $0.63$ m and a pixel size of $55~\mu$m~\cite{Cha2016}.
	}
\end{table}
 
Firstly, Table~\ref{tab:canabletoRecover} shows the number of pixel measurement constraints in our simulations as a function of the X-ray energy multiplier and the number of detector positions. 
The maximum number of pixel measurements at each PBF is computed using Equation~\eqref{eq:constraintsDiagonalOffset} of the Appendix.
The two regions of interest are: 
(\romannum{1}) the numbers coloured in red which fall below the theoretical limit of $M = K \log (N^2 / K)$, and
(\romannum{2}) the grey-shaded numbers which show that further detector offsets along the diagonals do not give additional binning constraints due to the grid periodicity.
Here, we aimed to recover the ground truth within a $120  \times 120$ pixel region,
such that  $N = 120$.
The number of unknown sub-pixels to recover is therefore $N^2 = 14400$.
Also, $K \simeq 1499$ for the  simulated crystal (\emph{i.e.} the number of significant components in the DCT of the central diffraction slice).
The numbers in  region (\romannum{1}) result from an inadequate number of detector positions. In this regime, the acquired signal information is truly deficient and sparse recovery is not possible.
Region (\romannum{2}) denotes the limitations of the chosen detector translation strategy (in this case, diagonal offsets only). 
For a given crystal size and detector pixel size, appropriate choices of sample-detector distance and detector offset strategy ensure the existence of the intermediate region between red and grey where compressed sensing can successfully recover the diffracted intensity pattern.

The connection between $M$ and the experimental parameters is made if we imagine that $M$ binning constraints can also be obtained from a \emph{single} detector image of $\sqrt{M} \times \sqrt{M}$ pixels that queries the same region of interest in reciprocal space. 
Stated differently, the set of offset detector images with a larger pixel size is conceptually equivalent to a single detector image with a smaller pixel size.
From an information theoretical point of view, the two sets of constraints are equivalent descriptions of the scattered wave intensity.
This immediately suggests a reinterpretation of Table~\ref{tab:canabletoRecover} in terms of experimental parameters.

For example, at a beam energy of $54$ keV, we have PBF = $6$ relative to the same measurement at 9 keV.
The original $ 120 \times 120$-pixel $q$-space region of interest in the ground truth simulation is now squeezed into a $20 \times 20$-pixel region in the centre of the detector due to reciprocal space compression.
The effective coarse pixel size over the original region of interest is simply $\Delta x = 1/20 = 0.05$ units.
But with $7$ detector positions at this energy, Table~\ref{tab:canabletoRecover} shows that we have access to a total of $2566$ pixel measurements, effectively emulating a $\sqrt{2566} \times \sqrt{2566}$ pixel grid over the same region of interest.
The effective pixel size is now scaled by a factor of $20/\sqrt{2566}$, or equivalently, we are able to resolve spatial frequencies better by a factor of $\sqrt{2566}/20 = 2.533$, which we label $f$. 

In the Fraunhofer approximation, the quantity $f$ can also be interpreted as an effective increase in sample-detector distance, given a fixed pixel size.
The spatial frequency step is related to the pixel size $\Delta x$ and sample-detector distance $z$ by $\Delta k \propto \Delta x / z$.
A scaled pixel size of $\Delta x / f$ implies an effective sample-detector distance of $fz$.
Table~\ref{tab:canabletoRecover-2} shows Table~\ref{tab:canabletoRecover} reinterpreted in terms of the distance multiplier $f$.

\begin{figure}
	\centering
	\includegraphics[width=0.4\hcolwidth]{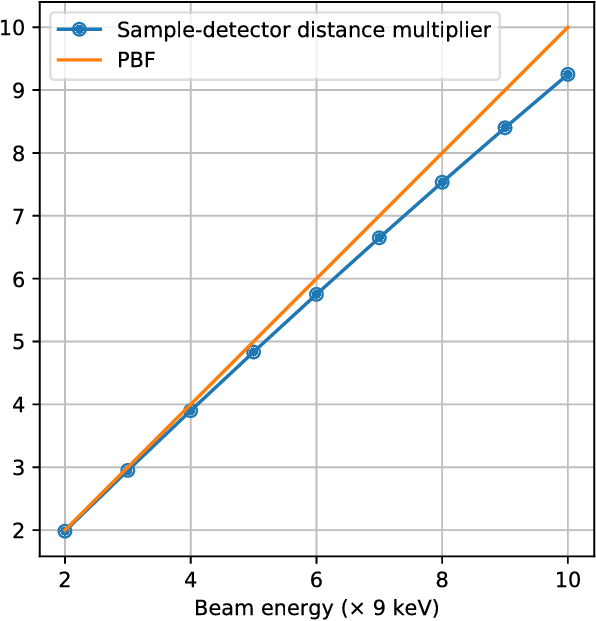}
	\caption{Behaviour of maximum value of sample-detector distance multiplier $f$ for different beam energies. For a fixed pixel size, the distance multiplier asymptotically approaches the PBF in the limit of large reciprocal space aperture (a wider detector).}
	\label{fig:scalingLimit}
\end{figure}

With a fixed beam energy, the limit of $f$ is reached when all possible binning constraints are accounted for and further detector offsets do not give new constraints. 
We have shown in Section~\ref{sup:underdetermined} of the Appendix that in this case, the number of pixel measurements that fall fully within the $q$-space region of interest is $M = \left[N + (\text{PBF} - 1)(N-1)\right]^2$.
Figure~\ref{fig:scalingLimit} shows the maximum values of $f$ under this condition, for a wider range of beam energies. 
We see that maximum upsampling nearly undoes the effect of pixel binning at high energies, through a proportionate increase of $f$, effectively emulating the sample-detector distance that would otherwise be needed to resolve intensity fringes. 
The difference is accounted for by the fact that pixels that are not fully contained within the original $q$-range are not used as measurement constraints.
This is closely related to the fact that the system of equations to be solved is inevitably underdetermined. 
 
\section{Summary}
\label{S:conclusion}
We have described a signal recovery technique for single-crystal BCDI data sets acquired at X-ray energies typically suited to HEDM applications ($>50$ keV). 
Our methodology relies on a modification of the conventional BCDI setup in which additional data is acquired by translating the detector across the incoming diffracted beam. 
We have described a fundamental incompleteness in the data collected by this method of upsampling, which calls for the incorporation of some form of extra information about the signal.
With this necessity we have motivated the use of techniques that recover not the signal directly, but sparse representations of it.
We have based our methodology on a rigorous proof of the existence of such a representation for all coherent scattering from a fully-illuminated compact single crystal.
We have quantified image recovery by this compressed sensing process and shown subsequent phase retrieval on the recovered data sets that are in agreement with the original simulated crystal.

Information theory- based analysis of detector-space upsampling shows us that smaller pixels and larger sample-detector distances can be emulated, with minimal changes to the experimental setup. 
This capability has the potential to significantly influence the design of space-constrained experiments at high energy coherent scattering beamlines.
When incorporated into existing near- and far-field HEDM workflows, high-energy BCDI could be the nanoscale component of a generalised experiment for imaging and characterisation of polycrystalline samples at multiple length scales.
Coupled with the development of new computational methodologies, such multiscale characterisation capabilities could go a long way in validating existing and new models of materials physics, as well as informing the creation and processing of engineering materials.
 
\section{Data availability}
The ground truth simulation and the Python code for sparse recovery are available upon reasonable request. Please contact the corresponding author.
 
\section{Competing financial interests}
The authors declare no competing financial interests.
 \section{Author contributions}
The analysis was developed by SM and implemented by SM with help from YN and ICA.
SM wrote the manuscript with contributions from SH, YN, RH and JSP.
All authors contributed to the refining of the concepts presented.
 
\section{Acknowledgements}
\label{sec:acknowledge}
Design, simulation, and demonstration of the in-plane detector translation upsampling framework for high-energy coherent x-ray diffraction was supported by Laboratory Directed Research and Development (LDRD) funding from Argonne National Laboratory, provided by the Director, Office of Science, of the U.S. Department of Energy under Contract No. DE-AC02-06CH11357.
Phase retrieval of the upsampled patterns and feasibility estimates were supported by the Advanced Photon Source, a U.S. Department of Energy (DOE) Office of Science User Facility operated for the DOE Office of Science by Argonne National Laboratory under Contract No. DE-AC02-06CH11357.
Adaptation of sparsity-based signal processing methods to coherent diffraction patterns upsampled via detector translation was supported by the U.S. Department of Energy, Office of Science, Basic Energy Sciences, Materials Science and Engineering Division.
 
\appendix
\section{Appendix}
\subsection{Detector-plane upsampling: an under-determined system of equations}
\label{sup:underdetermined}
Our goal is a method to reverse the binning process described in Section 2 of the main text and re-obtain the original high-resolution representation of a single detector image. 
\begin{figure}[H]
	\centering
	\subfloat[]{\includegraphics[width=0.6\hcolwidth]{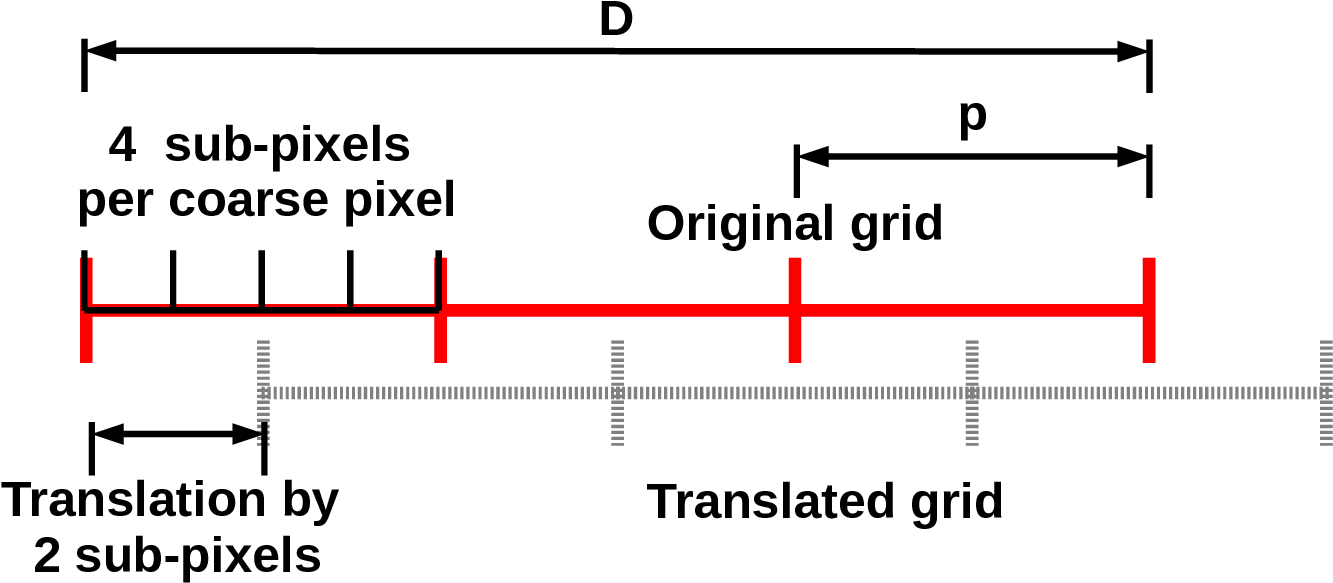}\label{fig:trans1}}	\hfill
	\subfloat[]{\includegraphics[width=0.35\hcolwidth]{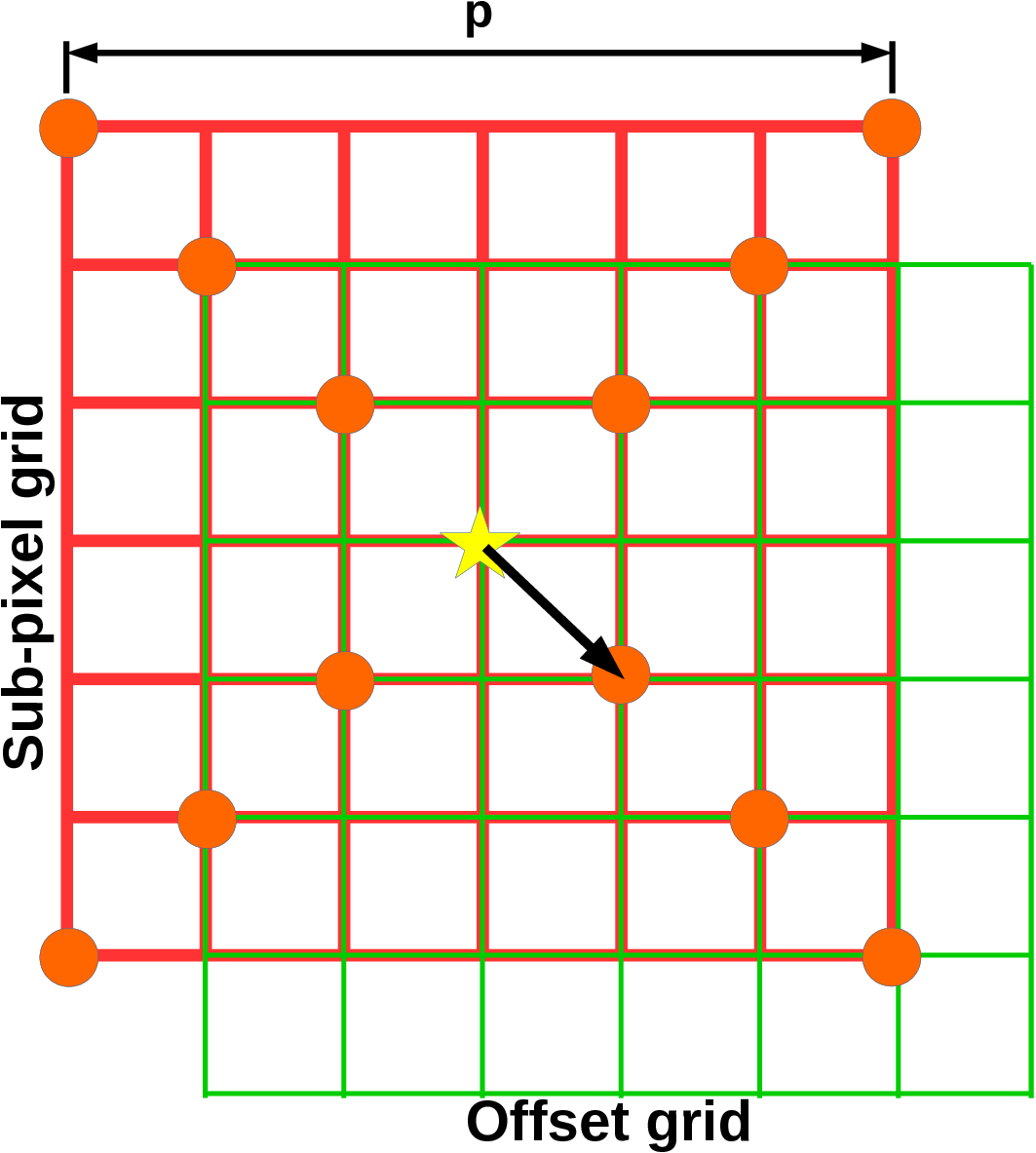}\label{fig:trans2}}
	\caption{
		\textbf{\protect\subref{fig:trans1}} Example of one-dimensional grid offsets in steps of a quarter of a physical pixel.
		\textbf{\protect\subref{fig:trans2}} Two-dimensional grid offsets in our simulations along the physical pixel diagonals. The star signifies the centre of the un-translated detector, while the orange markers denote the centres of the offset detector.
	}
	\label{fig:trans}
\end{figure}
In a high-energy BCDI experiment, the fine pixel size is determined by the smallest offset of the detector as a matter of convenience. 
Offsets in fractions of the chosen fine pixel size are undesirable since one is forced to make assumptions about the intensity distribution within a single fine pixel.
Under such sampling conditions, we can obtain constraints ( shown in the main text Figure 2) that couple the intensities of the fine "sub-pixels" with the measured coarse pixel intensities.
We show in this section that these sub-pixels always outnumber the constraints available through detector translation, regardless of the fineness of the offset.
In other words, the system of equations that couples the sub-pixels to coarse pixel measurements is always underdetermined, and has no unique solution. 
Some form of additional knowledge of the system is required to adequately constrain it.

To demonstrate the underdetermined nature of the problem, we consider one-dimensional binning for simplicity (refer to Figure~\ref{fig:trans1}).
Suppose we wish to upsample to a resolution of $m$ sub-pixels per coarse pixel. Here, $m$ is simply the pixel binning factor (PBF) of our high-energy simulations, and in our figure, $m = 4$.
If the original aperture size is $D$ with a pixel size $p$, the number of pixels is $N \equiv D / p$.
The periodicity of the binning grid implies that we need only offset the detector to $m-1$ sub-pixel positions to the right in order to generate unique constraints. 
Any further offset would simply result in redundant constraints. Any offset to the left can also be expressed as an equivalent offset to the right because of this periodicity.
Since we are only interested in the range of reciprocal space defined by the original aperture $D$ (red grid), we discard coarse pixel measurements that lie partially outside this region.
The total number of unique constraints is therefore: $M = N + (m-1)(N-1)$, where $N$ constraints come from the un-translated grid and each of the $m-1$ grid offsets contribute $N-1$ constraints.
On the other hand, the number of sub-pixels coupled by these constraints is $U = m \times N$. 

These numbers generalise to higher dimensions in a straightforward manner: $M_d = \left[N + (m-1)(N-1)\right]^d$ and $U_d = (mN)^d$ where $d$ is the dimensionality of the space.
The ratio of constraints to unknown quantities is therefore:
\begin{equation}
	\sigma_d = \frac{M_d}{U_d} = \left[1 - \frac{1}{N} + \frac{1}{mN}\right]^d < 1
\end{equation}
For two-dimensional images, $d = 2$.
Further, in the limit of infinitesimal detector offsets, $ m \rightarrow \infty$ and $\sigma_d \rightarrow (1-1/N)^d = (1-p/D)^d$, which is always  $< 1$.

Thus, a finite coarse pixel size $p$ ensures that $M_d$ is always less than $U_d$ and therefore the system of equations for the sub-pixels is always underdetermined and without a unique solution unless additional knowledge about the system is incorporated.
As an example, non-negativity could be enforced on the sub-pixel intensities.
In our methodology, the additional information is provided through the fact that the diffracted intensity pattern of a compact crystal has compact support in real space.
This physical insight applies to all compact single crystals and when successfully incorporated into the system of equations, paves the way for the use of sparse numerical solvers in high-energy BCDI. 

In our simulations, the detector was offset along the diagonals of the coarse pixels (Figure~\ref{fig:trans2}). 
Counting all possible constraints in this case is straightforward: 
if $m$ is odd, then there are $2m-1$ unique detector positions along the diagonals and if $m$ is even, there are $2m$ positions along the diagonals. 
In both cases, exactly one detector position (the zero-offset) contributes $N^2$ constraints, while the remaining contribute $(N-1)^2$ constraints. 
Thus the maximum number of unique constraints for diagonal offsets is given by:
\begin{equation}
	M = \left\{
		\begin{array}{cc}
			N^2 + ( 2m-1)(N-1)^2 & \text{ if }m\text{ is even} \\
			N^2 + ( 2m-2)(N-1)^2 & \text{ if }m\text{ is odd}
		\end{array}
		\right.
		\label{eq:constraintsDiagonalOffset}
\end{equation}
 
\subsection{Proof of signal sparsity}
\label{sup:appA}
The projection-slice theorem~\cite{Bracewell1956,Bracewell1990} states that any two-dimensional slice ($S$) of the three-dimensional Fourier transform ($F_3$) of a scalar field $f(\mathbf{x})$ can be alternately obtained by taking the two-dimensional Fourier transform ($F_2$) of the projection ($P_S$) of $f(\mathbf{x})$ in the slicing plane: 
\begin{equation}
	S \cdot F_3 f(\mathbf{x}) = F_2 \cdot P_S f(\mathbf{x})
	\label{eq:sliceprojection}
\end{equation}
If $f(\mathbf{x})$ is specifically the Patterson function of the scattering from a compact single crystal, then the LHS of Equation~\eqref{eq:sliceprojection} is the quantity that is measured on an area detector at each point in the rocking curve. 
Computing the 2D inverse Fourier transform throughout, we have: 
\begin{equation}
	F_2^{-1} \left[ S \cdot F_3 f(\mathbf{x}) \right] = F_2^{-1} F_2 P_S f(\mathbf{x}) = P_S f(\mathbf{x})
	\label{eq:sparseimageproof}
\end{equation}
The LHS of Equation~\eqref{eq:sparseimageproof} is the Fourier representation of a single detector image indexed by the slicing plane $S$, while the RHS is a planar projection of a compact Patterson function, and therefore is also compact.
The sparsity of the two-dimensional Fourier transform is ensured through provision of a sufficient buffer region as described in Section 3 of the main text.
Thus \emph{each detector image} in a Bragg CDI experiment has a sparse representation, making compressed sensing techniques applicable to each one independently.

\bibliography{MASTER}

\begin{thebibliography}{10}
\expandafter\ifx\csname url\endcsname\relax
  \def\url#1{\texttt{#1}}\fi
\expandafter\ifx\csname urlprefix\endcsname\relax\def\urlprefix{URL }\fi
\expandafter\ifx\csname doiprefix\endcsname\relax\def\doiprefix{DOI }\fi
\providecommand{\bibinfo}[2]{#2}
\providecommand{\eprint}[2][]{\url{#2}}

\bibitem{Robinson2001}
\bibinfo{author}{Robinson, I.~K.}, \bibinfo{author}{Vartanyants, I.~A.},
  \bibinfo{author}{Williams, G.~J.}, \bibinfo{author}{Pfeifer, M.~A.} \&
  \bibinfo{author}{Pitney, J.~A.}
\newblock \bibinfo{journal}{\bibinfo{title}{Reconstruction of the shapes of
  gold nanocrystals using coherent x-ray diffraction}}.
\newblock {\emph{\JournalTitle{Phys. Rev. Lett.}}}
  \textbf{\bibinfo{volume}{87}}, \bibinfo{pages}{195505}
  (\bibinfo{year}{2001}).
\newblock
  \urlprefix\url{https://link.aps.org/doi/10.1103/PhysRevLett.87.195505}.
\newblock \doiprefix 10.1103/PhysRevLett.87.195505.

\bibitem{Robinson2009}
\bibinfo{author}{Robinson, I.} \& \bibinfo{author}{Harder, R.}
\newblock \bibinfo{journal}{\bibinfo{title}{Coherent x-ray diffraction imaging
  of strain at the nanoscale}}.
\newblock {\emph{\JournalTitle{Nat Mater}}} \textbf{\bibinfo{volume}{8}},
  \bibinfo{pages}{291--298} (\bibinfo{year}{2009}).
\newblock \urlprefix\url{http://dx.doi.org/10.1038/nmat2400}.
\newblock \doiprefix 10.1038/nmat2400.

\bibitem{Ulvestad2015}
\bibinfo{author}{Ulvestad, A.} \emph{et~al.}
\newblock \bibinfo{journal}{\bibinfo{title}{Topological defect dynamics in
  operando battery nanoparticles}}.
\newblock {\emph{\JournalTitle{Science}}} \textbf{\bibinfo{volume}{348}},
  \bibinfo{pages}{1344--1347} (\bibinfo{year}{2015}).

\bibitem{Cha2016}
\bibinfo{author}{Cha, W.} \emph{et~al.}
\newblock \bibinfo{journal}{\bibinfo{title}{Three dimensional
  variable-wavelength x-ray bragg coherent diffraction imaging}}.
\newblock {\emph{\JournalTitle{Phys. Rev. Lett.}}}
  \textbf{\bibinfo{volume}{117}}, \bibinfo{pages}{225501}
  (\bibinfo{year}{2016}).
\newblock
  \urlprefix\url{http://link.aps.org/doi/10.1103/PhysRevLett.117.225501}.
\newblock \doiprefix 10.1103/PhysRevLett.117.225501.

\bibitem{Suter2008}
\bibinfo{author}{Suter, R.~M.} \emph{et~al.}
\newblock \bibinfo{journal}{\bibinfo{title}{Probing microstructure dynamics
  with x-ray diffraction microscopy}}.
\newblock {\emph{\JournalTitle{Journal of Engineering Materials and
  Technology}}} \textbf{\bibinfo{volume}{130}},
  \bibinfo{pages}{021007--021007--5} (\bibinfo{year}{2008}).
\newblock \urlprefix\url{http://dx.doi.org/10.1115/1.2840965}.
\newblock \doiprefix 10.1115/1.2840965.

\bibitem{Bernier2011}
\bibinfo{author}{Bernier, J.~V.}, \bibinfo{author}{Barton, N.~R.},
  \bibinfo{author}{Lienert, U.} \& \bibinfo{author}{Miller, M.~P.}
\newblock \bibinfo{journal}{\bibinfo{title}{Far-field high-energy diffraction
  microscopy: a tool for intergranular orientation and strain analysis}}.
\newblock {\emph{\JournalTitle{The Journal of Strain Analysis for Engineering
  Design}}} \textbf{\bibinfo{volume}{46}}, \bibinfo{pages}{527--547}
  (\bibinfo{year}{2011}).
\newblock \urlprefix\url{http://dx.doi.org/10.1177/0309324711405761}.
\newblock \doiprefix 10.1177/0309324711405761.
\newblock \eprint{http://dx.doi.org/10.1177/0309324711405761}.

\bibitem{Ludwig2008}
\bibinfo{author}{Ludwig, W.}, \bibinfo{author}{Schmidt, S.},
  \bibinfo{author}{Lauridsen, E.~M.} \& \bibinfo{author}{Poulsen, H.~F.}
\newblock \bibinfo{journal}{\bibinfo{title}{{X-ray diffraction contrast
  tomography: a novel technique for three-dimensional grain mapping of
  polycrystals. I. Direct beam case}}}.
\newblock {\emph{\JournalTitle{Journal of Applied Crystallography}}}
  \textbf{\bibinfo{volume}{41}}, \bibinfo{pages}{302--309}
  (\bibinfo{year}{2008}).
\newblock \urlprefix\url{https://doi.org/10.1107/S0021889808001684}.
\newblock \doiprefix 10.1107/S0021889808001684.

\bibitem{Liu2015}
\bibinfo{author}{Liu, Q.} \emph{et~al.}
\newblock \bibinfo{journal}{\bibinfo{title}{Quantifying the nucleation and
  growth kinetics of microwave nanochemistry enabled by in situ high-energy
  x-ray scattering}}.
\newblock {\emph{\JournalTitle{Nano letters}}} \textbf{\bibinfo{volume}{16}},
  \bibinfo{pages}{715--720} (\bibinfo{year}{2015}).

\bibitem{Chushkin2013}
\bibinfo{author}{Chushkin, Y.} \& \bibinfo{author}{Zontone, F.}
\newblock \bibinfo{journal}{\bibinfo{title}{{Upsampling speckle patterns for
  coherent X-ray diffraction imaging}}}.
\newblock {\emph{\JournalTitle{Journal of Applied Crystallography}}}
  \textbf{\bibinfo{volume}{46}}, \bibinfo{pages}{319--323}
  (\bibinfo{year}{2013}).
\newblock \urlprefix\url{https://doi.org/10.1107/S0021889813003117}.
\newblock \doiprefix 10.1107/S0021889813003117.

\bibitem{Candes2006}
\bibinfo{author}{Candes, E.~J.}, \bibinfo{author}{Romberg, J.} \&
  \bibinfo{author}{Tao, T.}
\newblock \bibinfo{journal}{\bibinfo{title}{Robust uncertainty principles:
  exact signal reconstruction from highly incomplete frequency information}}.
\newblock {\emph{\JournalTitle{IEEE Transactions on Information Theory}}}
  \textbf{\bibinfo{volume}{52}}, \bibinfo{pages}{489--509}
  (\bibinfo{year}{2006}).
\newblock \doiprefix 10.1109/TIT.2005.862083.

\bibitem{Donoho2006}
\bibinfo{author}{Donoho, D.~L.}
\newblock \bibinfo{journal}{\bibinfo{title}{Compressed sensing}}.
\newblock {\emph{\JournalTitle{IEEE Transactions on Information Theory}}}
  \textbf{\bibinfo{volume}{52}}, \bibinfo{pages}{1289--1306}
  (\bibinfo{year}{2006}).
\newblock \doiprefix 10.1109/TIT.2006.871582.

\bibitem{Goodman2005}
\bibinfo{author}{Goodman, J.}
\newblock \emph{\bibinfo{title}{Introduction to Fourier Optics}}.
\newblock McGraw-Hill physical and quantum electronics series
  (\bibinfo{publisher}{W. H. Freeman}, \bibinfo{year}{2005}).
\newblock \urlprefix\url{https://books.google.com/books?id=ow5xs\_Rtt9AC}.

\bibitem{Batey2014}
\bibinfo{author}{Batey, D.~J.} \emph{et~al.}
\newblock \bibinfo{journal}{\bibinfo{title}{Reciprocal-space up-sampling from
  real-space oversampling in x-ray ptychography}}.
\newblock {\emph{\JournalTitle{Phys. Rev. A}}} \textbf{\bibinfo{volume}{89}},
  \bibinfo{pages}{043812} (\bibinfo{year}{2014}).
\newblock \urlprefix\url{http://link.aps.org/doi/10.1103/PhysRevA.89.043812}.
\newblock \doiprefix 10.1103/PhysRevA.89.043812.

\bibitem{Makhoul1980}
\bibinfo{author}{Makhoul, J.}
\newblock \bibinfo{journal}{\bibinfo{title}{A fast cosine transform in one and
  two dimensions}}.
\newblock {\emph{\JournalTitle{IEEE Transactions on Acoustics, Speech, and
  Signal Processing}}} \textbf{\bibinfo{volume}{28}}, \bibinfo{pages}{27--34}
  (\bibinfo{year}{1980}).
\newblock \doiprefix 10.1109/TASSP.1980.1163351.

\bibitem{Wallace1991}
\bibinfo{author}{Wallace, G.~K.}
\newblock \bibinfo{journal}{\bibinfo{title}{The jpeg still picture compression
  standard}}.
\newblock {\emph{\JournalTitle{Commun. ACM}}} \textbf{\bibinfo{volume}{34}},
  \bibinfo{pages}{30--44} (\bibinfo{year}{1991}).
\newblock \urlprefix\url{http://doi.acm.org/10.1145/103085.103089}.
\newblock \doiprefix 10.1145/103085.103089.

\bibitem{Musmann2006}
\bibinfo{author}{Musmann, H.~G.}
\newblock \bibinfo{journal}{\bibinfo{title}{Genesis of the mp3 audio coding
  standard}}.
\newblock {\emph{\JournalTitle{IEEE Transactions on Consumer Electronics}}}
  \textbf{\bibinfo{volume}{52}}, \bibinfo{pages}{1043--1049}
  (\bibinfo{year}{2006}).
\newblock \doiprefix 10.1109/TCE.2006.1706505.

\bibitem{Shannon1949}
\bibinfo{author}{Shannon, C.~E.}
\newblock \bibinfo{journal}{\bibinfo{title}{Communication in the presence of
  noise}}.
\newblock {\emph{\JournalTitle{Proceedings of the IRE}}}
  \textbf{\bibinfo{volume}{37}}, \bibinfo{pages}{10--21}
  (\bibinfo{year}{1949}).
\newblock \doiprefix 10.1109/JRPROC.1949.232969.

\bibitem{Suykens2000}
\bibinfo{author}{Suykens, J.~A.}, \bibinfo{author}{Lukas, L.} \&
  \bibinfo{author}{Vandewalle, J.}
\newblock \bibinfo{title}{Sparse approximation using least squares support
  vector machines}.
\newblock In \emph{\bibinfo{booktitle}{Circuits and Systems, 2000. Proceedings.
  ISCAS 2000 Geneva. The 2000 IEEE International Symposium on}},
  vol.~\bibinfo{volume}{2}, \bibinfo{pages}{757--760}
  (\bibinfo{organization}{IEEE}, \bibinfo{year}{2000}).

\bibitem{Huang2008}
\bibinfo{author}{Huang, J.}, \bibinfo{author}{Ma, S.} \&
  \bibinfo{author}{Zhang, C.-H.}
\newblock \bibinfo{journal}{\bibinfo{title}{Adaptive lasso for sparse
  high-dimensional regression models}}.
\newblock {\emph{\JournalTitle{Statistica Sinica}}} \bibinfo{pages}{1603--1618}
  (\bibinfo{year}{2008}).

\bibitem{Candes2011}
\bibinfo{author}{Candès, E.~J.}, \bibinfo{author}{Eldar, Y.~C.},
  \bibinfo{author}{Needell, D.} \& \bibinfo{author}{Randall, P.}
\newblock \bibinfo{journal}{\bibinfo{title}{Compressed sensing with coherent
  and redundant dictionaries}}.
\newblock {\emph{\JournalTitle{Applied and Computational Harmonic Analysis}}}
  \textbf{\bibinfo{volume}{31}}, \bibinfo{pages}{59 -- 73}
  (\bibinfo{year}{2011}).
\newblock
  \urlprefix\url{http://www.sciencedirect.com/science/article/pii/S1063520310001156}.
\newblock \doiprefix http://dx.doi.org/10.1016/j.acha.2010.10.002.

\bibitem{Tibshirani1996}
\bibinfo{author}{Tibshirani, R.}
\newblock \bibinfo{journal}{\bibinfo{title}{Regression shrinkage and selection
  via the lasso}}.
\newblock {\emph{\JournalTitle{Journal of the Royal Statistical Society. Series
  B (Methodological)}}} \bibinfo{pages}{267--288} (\bibinfo{year}{1996}).

\bibitem{Chapman2006}
\bibinfo{author}{Chapman, H.~N.} \emph{et~al.}
\newblock \bibinfo{journal}{\bibinfo{title}{High-resolution ab initio
  three-dimensional x-ray diffraction microscopy}}.
\newblock {\emph{\JournalTitle{J. Opt. Soc. Am. A}}}
  \textbf{\bibinfo{volume}{23}}, \bibinfo{pages}{1179--1200}
  (\bibinfo{year}{2006}).
\newblock
  \urlprefix\url{http://josaa.osa.org/abstract.cfm?URI=josaa-23-5-1179}.
\newblock \doiprefix 10.1364/JOSAA.23.001179.

\bibitem{Marchesini2007}
\bibinfo{author}{Marchesini, S.}
\newblock \bibinfo{journal}{\bibinfo{title}{Invited article: A unified
  evaluation of iterative projection algorithms for phase retrieval}}.
\newblock {\emph{\JournalTitle{Review of Scientific Instruments}}}
  \textbf{\bibinfo{volume}{78}}, \bibinfo{pages}{011301}
  (\bibinfo{year}{2007}).
\newblock \urlprefix\url{http://dx.doi.org/10.1063/1.2403783}.
\newblock \doiprefix 10.1063/1.2403783.
\newblock \eprint{http://dx.doi.org/10.1063/1.2403783}.

\bibitem{Bracewell1956}
\bibinfo{author}{Bracewell, R.~N.}
\newblock \bibinfo{journal}{\bibinfo{title}{Strip integration in radio
  astronomy}}.
\newblock {\emph{\JournalTitle{Australian Journal of Physics}}}
  \textbf{\bibinfo{volume}{9}}, \bibinfo{pages}{198--217}
  (\bibinfo{year}{1956}).
\newblock \urlprefix\url{https://doi.org/10.1071/PH560198}.

\bibitem{Bracewell1990}
\bibinfo{author}{Bracewell, R.~N.}
\newblock \bibinfo{journal}{\bibinfo{title}{Numerical transforms}}.
\newblock {\emph{\JournalTitle{Science}}} \textbf{\bibinfo{volume}{248}},
  \bibinfo{pages}{697--704} (\bibinfo{year}{1990}).
\newblock \urlprefix\url{http://science.sciencemag.org/content/248/4956/697}.
\newblock \doiprefix 10.1126/science.248.4956.697.
\newblock \eprint{http://science.sciencemag.org/content/248/4956/697.full.pdf}.

\end{thebibliography}
 
\end{document}